\documentclass[a4paper,11pt]{article}
\usepackage{pos}
\usepackage{amsmath}
\usepackage{cleveref}
\usepackage{gensymb}
\usepackage{colortbl}

\usepackage{enumerate}
\usepackage{subcaption}

\newcommand{\lp}{\left(}
\newcommand{\rp}{\right)}

\newcommand{\TeV}{\text{TeV}}
\newcommand{\GeV}{\text{GeV}}

\newcommand{\beq}{\begin{equation} }
\newcommand{\eeq}{\end{equation}} 
\newcommand{\bi}{\begin{itemize} }
\newcommand{\ei}{\end{itemize} }

\newcommand{\g}{\gamma }

\definecolor{nicered}{rgb}{0.7,0.1,0.1}
\definecolor{nicegreen}{rgb}{0.1,0.5,0.1}
\definecolor{violet}{rgb}{0.7,0.3,0.3}
\hypersetup{colorlinks,citecolor= nicegreen,linkcolor= nicered}

\title{Axion searches at colliders}

\author[a]{Michele Tammaro}
\author[b]{Jure Zupan}

\affiliation[a]{INFN Sezione di Firenze, Via G. Sansone 1, I-50019 Sesto Fiorentino, Italy}
\affiliation[b]{Department of Physics, University of Cincinnati, Cincinnati, Ohio 45221, USA}

\emailAdd{michele.tammaro@fi.infn.it}
\emailAdd{zupanje@ucmail.uc.edu}

\abstract{We provide an introduction to searches for axions and Axion-Like Particles (ALPs) at colliders. After covering the basics of collider physics, with a focus on production and detection of new particles, we give a rather broad introduction into searches for dark sectors at colliders. While our focus is on searches for light dark sectors, with the special emphasis on ALPs, we also contrast these with searches for heavy dark matter. 
In the final part of the notes we provide a ``tutorial'', a worked out example of  a search for a photo-philic ALP  in $e^+e^-$ collisions, reflecting the actual analysis that was performed at Belle-II with early data. }

\FullConference{2nd Training School and General Meeting of the COST Action COSMIC WISPers  (CA21106) (COSMICWISPers2024)\\
 10-14 June 2024 and 3-6 September 2024\\
Ljubljana (Slovenia) and Istanbul (Turkey)\\}

 \tableofcontents

\begin{document}
\maketitle

\section{Introduction}
Axions, and more generally, Axion-Like-Particles (ALPs) are increasingly popular candidates for dark matter, or for constituents of dark sectors that may accompany dark matter. In this lecture notes we review how such dark sectors can be searched for at colliders. To get us going we will first review the basics of collider physics, which will then equip us to understand what types of signatures dark sectors could lead to in collider experiments. Our primary focus will be on searches for QCD axion and ALPs, using colliders.  While we will keep the discussion quite short, there are a number of excellent more detailed reviews both on the topic of dark matter in general (see, e.g., \cite{Cirelli:2024ssz,Bertone:2010zza,Bertone:2004pz}), as well as on axion dark matter specifically (see, e.g., \cite{OHare:2024nmr,DiLuzio:2020wdo}),  that we refer the reader to for further study. 

Before we move on to the main content of the lectures a small note on terminology. Throughout the lecture notes when we refer to {\em ``axions''} we really mean a {\em ``QCD axion''}, a pseudo-Nambu-Goldstone boson (pNGB) whose mass comes from coupling to QCD anomaly. That is, the Lagrangian for a light pseudoscalar $a$ that couples to gluons through a higher dimension operator is given by
\beq
\label{eq:L:axion}
{\cal L}=\frac{1}{2} \big(\partial_\mu a\big)^2-\frac{1}{2} m_a^2 a^2 -\frac{1}{f_a} \frac{\alpha_s}{8\pi} G_{\mu\nu}^a \tilde G^{a\mu\nu},
\eeq
where a summation over repeated indices is understood, $G_{\mu\nu}^a \tilde G^{a\mu\nu}=\frac{1}{2}\epsilon^{\mu\nu\gamma\delta} G^a_{\gamma\delta}$, and the axion decay constant $f_a$ is a dimensionful parameter, comparable to the UV scale at which the global symmetry giving rise to pNGB is spontaneously broken. For QCD axion, the mass is zero in the UV, i.e., $m_a=0$ in \eqref{eq:L:axion}. This does not mean that axion remains massless in the IR. Below the QCD confinement scale the interactions with gluons create a potential for $a$ that results in a tiny mass \cite{GrillidiCortona:2015jxo},
\beq
\label{eq:QCD:axion:mass}
m_a=\frac{m_\pi f_\pi}{f_a}=0.57 \text{\,meV} \times \biggr(\frac{10^{10}\text{\,GeV}}{f_a}\biggr)=0.57\text{\,keV} \times \biggr(\frac{10\text{\,TeV}}{f_a}\biggr),
\eeq
where $m_\pi$ is the pion mass, $f_\pi$ the pion decay constant, and we used two representative values for $f_a$ in the two numerical examples above. As we see, the mass $m_a$ is so small that the QCD axion always appears as effectively massless particle at colliders.

This in contrast to the case of {\em axion-like particles} (ALPs), for which the mass $m_a$ is treated as a free parameter. As a result, ALPs can be significantly heavier than QCD axions, with masses extending into the MeV or even GeV range. At collider experiments, they would therefore appear as massive particles. This distinction between axions and ALPs has important implications for collider phenomenology, as we will explore below. Before delving into these details, however, we will first briefly review the basic principles of high-energy collider experiments.

The lecture notes are thus organized as follows. In \cref{sec:BasicsOfCollider}, we begin with a brief overview of collider physics. \cref{sec:search:dark:sector} introduces the main strategies for searching for dark sectors at colliders, while \cref{sec:models:dark:forces} focuses on the physics of light dark sectors, including their production and decays, with particular emphasis on axions and ALPs. The final part of these notes, presented in \cref{sec:tutorial}, consists of a hands-on tutorial.

\section{Basics of collider physics}
\label{sec:BasicsOfCollider}
What can and cannot be measured in high-energy experiments is determined by how these experiments are designed. There are two key factors: the mechanisms by which particles are produced, and how they are detected. 
In practice, one often faces trade-offs between precision, feasibility, and cost. 
As a result, detectors are typically optimized for specific tasks or measurements, realizing in practice the well-known quote attributed to Albert Einstein, ``Everything should be made as simple as possible, but not simpler''. 
In what follows, we briefly review the advantages and limitations of different experimental approaches, partly drawing from the summary in~\cite{Cirelli:2024ssz}.

\subsection{Production}
The production of particles can be a result of two beams colliding, such as in electron-positron colliders and in hadron colliders. The new particles can also be produced by colliding beam with a target at rest in the so called ``fixed-target experiments''. 
\begin{description}
\item {\bf Electron-positron colliders}. To reach high energies, electrons and positrons are accelerated on circular orbits and then collided at one or several collision points. The energy of such $e^+e^-$ collider is limited due to the energy losses via Larmor radiation, where the energy loss per path traveled is given by $dE/dx = 2 \alpha_{\rm em} \gamma^4\beta^3/3{R^2}$, with $\gamma =E/m_e$ the Lorentz factor,  $\alpha_{\rm em}\simeq 1/137$ the fine structure constant, $\beta$ the velocity of the electron, and $R$ the radius of the circular orbit. The largest possible energy, $E_{\rm max}=\gamma_{\rm max} m_e$ is 
set by $dE/dx|_{\rm max}$, the largest possible energy per linear section of the orbit  that can be delivered to the accelerating $e^\pm$, giving (we set $\beta\simeq 1$)
\beq 
\frac{E_{\rm max}}{m_e}\sim 6\cdot10^5\biggr({\frac{R}{5\,\text{km}}}\biggr)^{1/2}\biggr({\frac{dE/dx|_{\rm max}}{m_ec^2/10\,\text{cm}}}\biggr)^{1/4}.
\eeq
The most energetic collider of this type was LEP, which reached the center-of-mass energy $\sqrt{s}\approx 200\,\GeV$ and had a radius $R\approx 4.3$\,km. The future versions of such a collider, FCC-ee \cite{Blondel:2021ema,FCC:2018evy} and CEPC \cite{CEPCStudyGroup:2018ghi}, with $R\approx 15$\,km and thus a higher energy, are being discussed. Note, however, that the increase in energy is only $\propto R^{1/2}$, keeping everything else the same.
The alternative is to build linear $e^+e^-$ colliders, for which  the maximal energy is limited by their length $L$,  
\beq 
E_{\rm max} = L   \left.\frac{dE}{dx}\right|_{\rm max}  = 1\TeV \frac{L}{10\text{km}} \frac{dE/dx|_{\rm max}}{1 \text{MeV}/{\rm cm}}.
\eeq
Circular muon colliders could reach much higher energies \cite{Accettura:2023ked}, since due to $m_\mu/ m_e\approx 210\gg 1$ the synchrotron losses are much smaller. A number of technical challenges, such as how to efficiently accelerate muons before they decay, need to be solved first, though. 

\item  {\bf Hadron colliders.} The most energetic examples are the currently running LHC ($pp$ collisions at $\sqrt{s}=13.6\TeV$) and the already finished Tevatron ($p\bar p$ collisions at $\sqrt{s}\sim 2\TeV$), which stopped taking data  in  2011. In hadron colliders the Larmor radiation is negligible, since $m_p\gg m_e$. The limiting factor is instead the strength of the magnetic field $B$, needed to have charged particles of energy $E$ moving on circular orbits of radius $R$. Neglecting proton mass, we have
\beq  
E=QRB = 15\TeV \frac{Q}{e}\frac{R}{5{\rm km}}\frac{B}{10\,{\rm Tesla}}. 
\eeq
Magnetic fields exceeding tens of Tesla cause mechanical failures, so that bigger energies require larger colliders. The proposed future $pp$ collider, FCC-hh, with $\sqrt{s}\sim 100\,\TeV$, would have $R\sim 15$\,km (and would thus fit in the FCC-ee tunnel, in the same way as the LHC replaced LEP). Performing particle physics using hadron colliders poses several unique challenges. First of all, unlike electrons, protons are composite and thus their constituents, quarks and gluons, carry only a variable fraction $x\sim 0.1$ of the total proton momentum. Each elementary collision thus occurs at a reduced energy $\sqrt{\hat{s}} =\sqrt{ x_1 x_2 s}$ that is about an order of magnitude smaller than $\sqrt{s}$. Furthermore, $pp$ collisions occur with a large QCD cross-section, $\sigma(pp)\sim 1/m_\pi^2$, where $m_\pi=0.14$ GeV is the pion mass. In $pp$ collisions there are thus large QCD backgrounds to the elementary
processes that one is interested in, such as the production of ALPs.

\item  {\bf Fixed target experiments.} 
 Fixed target experiments are simpler than the hadron and electron colliders, since one needs to only create one beam, which then collides with the target at rest. Due to its relative simplicity the beam can be made out of protons or electrons, but also out of muons. Another benefit is that beams can be optimized for intensity, so that one can collect large data samples. On the negative side, the center of mass of the collision, $\sqrt{s}=\sqrt{2 E m_p}$, is much smaller than the energy of the beam $E$ (here $m_p$ is the mass of the proton as a stand-in for the mass of the nucleons in the target material, and we neglected the mass of the particles in the beam). Especially interesting for ALP searches  are the so called {\em beam dump} experiments, in which the beam is dumped into a dense block of a heavy material, so that the hadronic cascade is absorbed as fast as possible.
\end{description}

\subsection{Detection}

\begin{figure}[t]
	\centering
    \includegraphics[width=.6\linewidth]{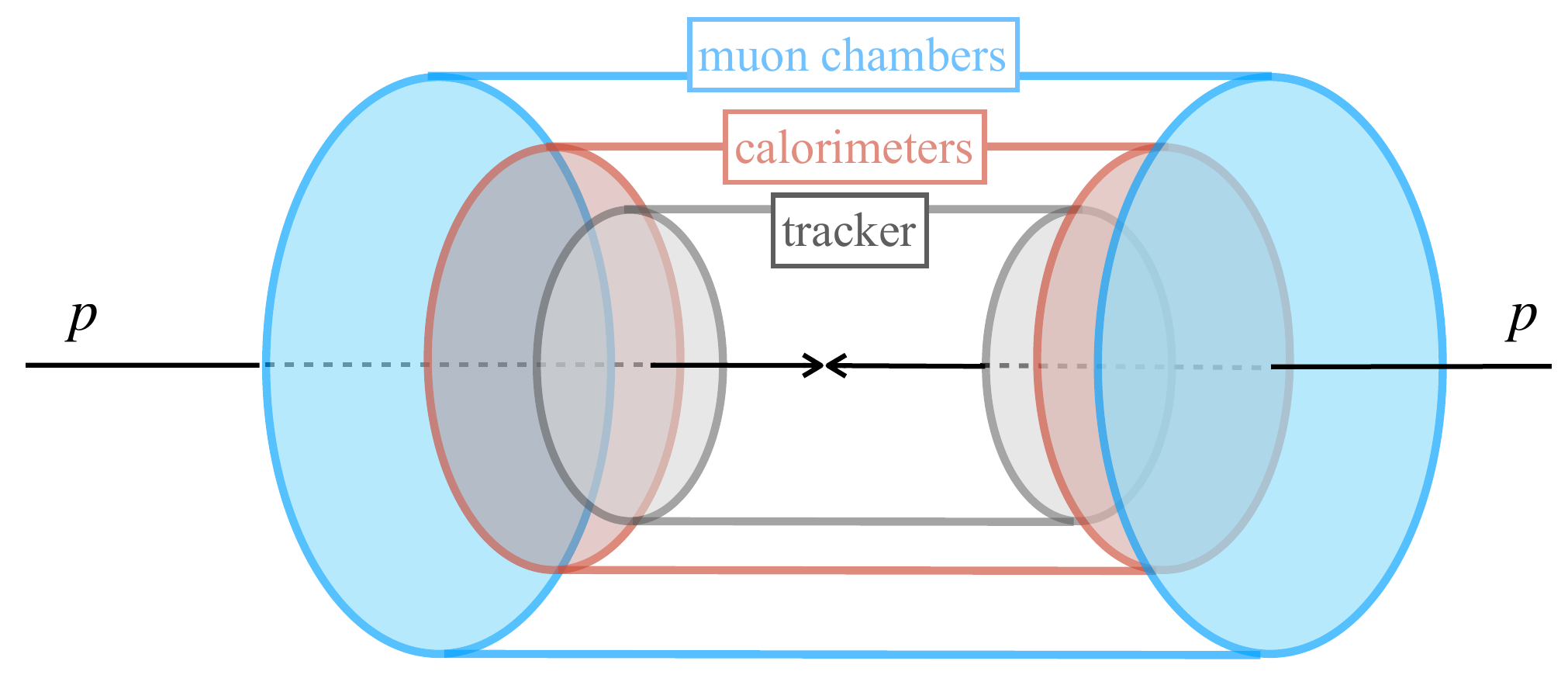}
    \caption{\em A sketch of the onion like structure of a multi-purpose particle physics detector used in $pp$ and $e^+e^-$ collisions, with tracking components closest to the interaction point, followed by the electromagnetic and hadron calorimeters and finally the muon chambers.  }
\label{fig:detector}
\end{figure}

A typical multi-purpose particle physics detector surrounds collision in an onion-layered design, see \cref{fig:detector}, where each detector subcomponents serves  a dedicated purposes.  Closest to the interaction point are vertex detectors, a highly precise tracking device used to pinpoint the precise location where particles interact or decay. Further out are calorimeters that measure energy that is deposited either via electromagnetic or hadronic showers, while furthest removed from the interaction point are the muon chambers. Examples of such experiments are ATLAS \cite{ATLAS:2008xda}, CMS \cite{CMS:2008xjf} at the LHC, for $pp$ collisions, and Belle 2 \cite{Belle-II:2010dht} at KEK for $e^+e^-$ collisions.

\begin{figure}[t]
	\centering
    \includegraphics[width=.8\linewidth]{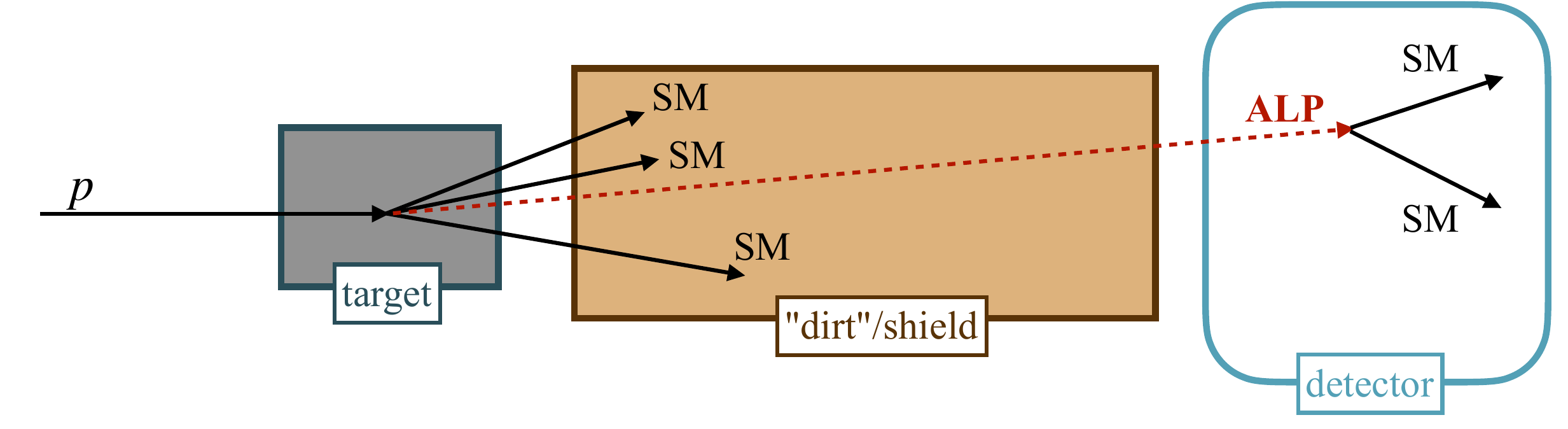}
    \caption{\em A schematics of a proton beam dump experiment, creating standard model (SM) particles and the ALP. SM particles get stopped in the shield, while a feebly interacting ALP can make it all the way to the detector, where it can be detected via its decays to the SM particles. }
\label{fig:beam:dump}
\end{figure}

In beam dumps the target is followed by a shield and a detector placed behind it, see \cref{fig:beam:dump}. The shield can be very long, composed of a few 100\,m of dirt. This configuration allows to search for long lived particles that are created in collisions of a beam with the target, but then interact very feebly, traversing the shield, and decaying inside the detector. Examples are the planned Codex-b experiment near the LHCb collision point at the LHC \cite{CODEX-b:2019jve,CODEX-b:2024tdl,Gligorov:2017nwh}, and an already operating experiment, FASER \cite{Feng:2017uoz}, which is $\sim 500$\,m away from ATLAS collisions point and lies on the beam collision axis line of sight, but hidden behind dirt after the LHC ring starts to curve. 

It is important to remember that we can detect directly only the  lightest few particles, which are either stable or long-lived on collider timescales, such that they leave visible imprints:

\begin{itemize}
\item[$\circ$] {\bf Electrons and photons} are seen in electro-magnetic calorimeters. These devices can reliably identify them and measure their
energies with $\approx 1\%$ level precision.

\item[$\circ$] {\bf Muons} appear as almost-stable minimally ionizing particles that reach the outer layers of detectors. 

\item[$\circ$] {\bf Quarks and gluons} are bound inside hadrons: $\pi^\pm$, $p$, $n$, $K$,\ldots. At collision energies well above the QCD confinement scale of ${\mathcal O}(1\,\GeV)$ the creation of quarks and gluons in elementary collisions ({\em the hard process}) appears as sprays of hadrons roughly collimated in a particular direction, the so called {\em jets }. Their energy is measured in hadronic calorimeters with $\approx 10\%$ uncertainty.
A tracker positioned near the production point can identify bottom quarks (and to some extent charm quarks) based on their displaced decays.

\item[$\circ$] {\bf Neutrinos} interact very feebly, and escape from detectors without interacting with the detector elements. We can infer their production only indirectly, by measuring all the other visible particles, and then deducing that a fraction of the total energy and momentum is {\em missing}, since it was carried away by neutrinos. 

\end{itemize}
Heavier particles can be identified via their decay products. For example, a $Z$ boson decaying
into a $\mu^+\mu^-$ pair will appear as a resonant peak at $Z$ boson mass in the cross section for $e^+e^-\to \mu^+\mu^-$ as a function of invariant mass $M_{\rm inv}^2\equiv(P_{\mu^+} + P_{\mu^-})^2$ , where $P_{\mu^\pm}$ are the muon four-momenta. Similarly, one can identify long-lived particles (LLPs) from their decays. Due to their long lifetimes, they would propagate a long distance before decaying in the detector.  For LLP signal the sum of the decay particles' momenta should point away from the beam dump target, while the presence of the shield should get rid of most of the backgrounds that could be confused with the signal. 

There are a number of choices that can be made regarding the detectors, that may give more information about the particles being produced. For instance, if the detector is inside a magnetic field, this then allows to distinguish $e^+$ from $e^-$, and similarly $\mu^+$ from $\mu^-$, as well as to measure their momenta from the curvatures of their trajectories. There are also detector elements that can be used to distinguish pions from kaons and these from proton, for instance, from their losses in the detectors. All these additional handles can be useful in identifying the underlying fundamental physical processes.

\section{Searching for dark sectors at colliders}
\label{sec:search:dark:sector}
In this section we will first review general searches for dark matter and dark sectors at colliders, and then specialize in \cref{sec:models:dark:forces} to ALP/axions searches, highlighting the differences between the general DM and ALP searches.

\subsection{Heavy dark matter}
The searches for heavy dark matter (DM) rely on the fact that DM needs to be uncharged under QCD and QED, and needs to be stable. ``Heavy'' for us means that it is roughly of the electroweak scale, i.e., with a mass in the $\sim 100$\,GeV to few TeV range. Traditionally, such heavy DM mass was denoted as a WIMP (weakly interacting massive particle). Strictly speaking, the name WIMP also implies that the DM particle carries an electroweak charge, however, whether or not that is the case, is not important for our purposes (see \cite{Cirelli:2024ssz} for a more detailed discussion of WIMPs).

\begin{figure}[t]
	\centering
    \includegraphics[width=.27\linewidth]{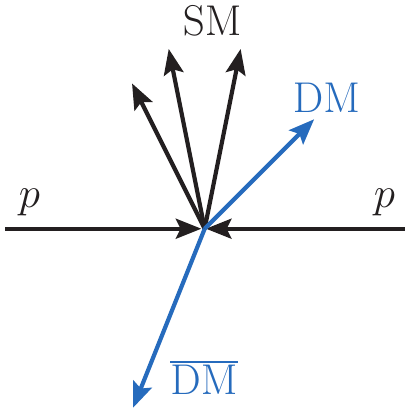}~~~~~
   \includegraphics[width=.27\linewidth]{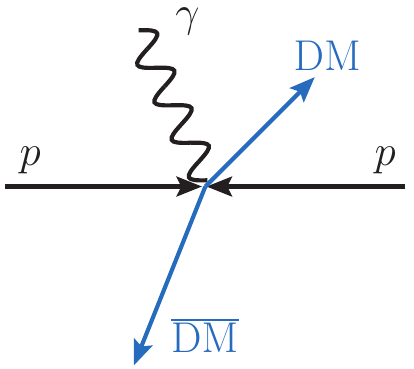}~~~~~
      \includegraphics[width=.27\linewidth]{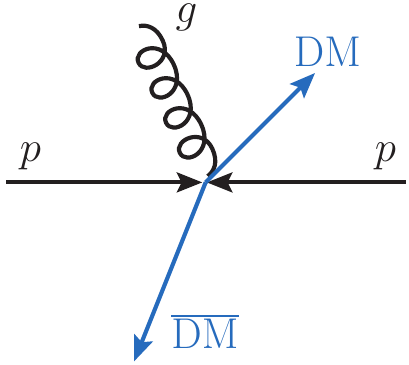}
    \caption{\em DM pair production in $pp$ collisions, where the examples of SM particles in the left bin can be a single hard photon (middle panel) or hard gluon (right panel). The single hard gluon hadronizes into a spray of SM hadrons (pions, kaons, etc) --- a  jet.   }
\label{fig:heavy:DM:pair:prod}
\end{figure}

Heavy DM cannot be singly produced at colliders; a single production of heavy DM is forbidden by the requirement that it needs to be stable on cosmological timescales. That is, if the single production $pp\to\text{DM}+\text{SM}$ is possible then also the reverse processes, with the SM in the initial state, are allowed, including the DM decay,  $\text{DM}\to pp+\overline{\text{SM}}$ (here ``SM'' denotes a collection of stable or almost stable SM particles, such as $e,\mu, \pi, K, p$, all of which have masses much smaller than DM). A simple symmetry that can ensure the stability of DM is a $Z_2$ under which DM is odd, i.e, under $Z_2$ dark matter transforms as $\text{DM}\to -\text{DM}$, while the SM particles are even.  Heavy DM can then be produced  at a collider only in pairs, such as (see also \cref{fig:heavy:DM:pair:prod})
\beq
\label{eq:DM:pair:prod}
pp\to \text{DM}+{\text{DM}}, \qquad \text{or} \qquad pp\to \text{DM}+\overline{\text{DM}},
\eeq
where in the first case DM is its own antiparticle. (More complicated theories may even require that production of dark matter in the collider is only possible if at least three DM particles are produced at the same time.) Pair production, \cref{eq:DM:pair:prod}, means that in order to produce DM collider energy needs to be at least 
\beq
\sqrt{\hat s}>2M_\text{DM},
\eeq
where $M_\text{DM}$ is the DM mass. 
Once produced, DM particles escape the detector. Production of DM therefore results in a missing energy signature, in the same way as the production of the SM neutrinos. The production of SM neutrinos in the collisions is thus an irreducible background to DM searches. 

Furthermore, missing energy can only be observed if there are additional visible particles that are produced in the collisions, so that the two DM particles recoils agains something ``visible'' that can be detected. Important examples are the {\em monophoton} signature: a collision that results in an additional photon, $pp\to \text{DM}+\overline{\text{DM}}+\gamma$; and the {\em monojet} signature: a production of and additional hadronic jet, $p p\to \text{DM}+\overline{\text{DM}}+j$. At the partonic level, this would arise for instance from quark--anti-quark annihilation, with the emission of a gluon, $q\bar q\to  \text{DM}+\overline{\text{DM}}+g$.

\subsubsection{Example: searching for SUSY DM}
Historically, a very important example of a heavy DM was supersymmetric (SUSY) DM (for excellent introduction, see, e.g., Ref. \cite{Martin:1997ns}). The Higgs hierarchy problem, i.e., why is Higgs so much lighter than the Planck mass, motivated introduction of a new supersymmetric partner for each SM particle. If (at least some of)  the SUSY partners had masses below about 1 TeV this would then ensure that  the otherwise quadratically divergent radiative corrections to the Higgs mass would remain reasonably small. The lightest SUSY particle (LSP, probably a {\em neutralino} $\chi_0$) is a good DM candidate that could be produced in the early Universe from the so called freeze-out mechanism in correct quantities to explain the observed DM relic abundance. 

Before the LHC started operating, the $pp$ collisions were expected to produce large numbers of SUSY particles that are charge under the strong interactions: gluinos $\tilde g$ (the SUSY partners of gluons), and squarks $\tilde g$ (the SUSY partners of quarks). These would ultimately decay into DM. These decay chains could be quite long, for instance, 
\beq
pp \to \tilde g\tilde g, \qquad \text{where}\qquad \tilde g\to \bar q \tilde q\to \bar q q \ell \nu \chi_0.
\eeq
These long decay chains  would lead to a miriad of searchable signatures, that are much more specific than simple generic searches for DM that rely only on monophoton or mono-jet signatures. Quite importantly, since DM would be produced from decays of colored particles, the cross section for production of DM would be quite large, given by the QCD cross sections for the productions of colored states, gluinos  and/or squarks. While no such spectacular signals with many particles were observe at the LHC so far, with about 1/30th of the expected total integrated luminosity, the above example does teach us an important lesson. Even though DM itself is not charged under strong interactions, and thus interacts only feebly with the ordinary matter, it can still be produced with relatively large rates in $pp$ collisions, if it is accompanied by additional colored states. 

\subsubsection{Example: colored co-annihilation}
An interesting scenario of this type  that is particularly relevant for the LHC, and hadron colliders in general, is the case where DM is accompanied by a slightly heavier colored particle $\text{DM}'$ \cite{Baer:1998pg}. In the SUSY example above, $\text{DM}'$ could be $\tilde g$ or $\tilde q$. Since the colored particle is not much heavier than DM then, even if $\text{DM}+\text{DM}$ annihilations are negligible, the $\text{DM}+\text{DM}'$ annihilations in the early Universe can still result in a correct DM relic abundance. That is, even if $\text{DM}+\text{DM}\to \text{SM}+\text{SM}$ annihilation cross section is very small, the correct relic abundance can still be produced via the co-annihilation process where DM first converts to $\text{DM}'$ via $\text{DM}+\text{SM}'\to \text{DM}'+\text{SM}$, and then $\text{DM}'$ efficiently annihilates away,  $\text{DM}'+\text{DM}'\to \text{SM}+\text{SM}$. Since $\text{DM}'$ is a colored particle, the $\text{DM}'+\text{DM}'\to \text{SM}+\text{SM}$ annihilations proceed via a large, QCD sized cross section
\beq
\label{eq:DM':annih}
\sigma(\text{DM}'+\text{DM}'\to gg+q\bar q)  v_{\rm rel}=\frac{\pi \alpha_s^2}{M_{\rm DM}^2} {\mathcal O}(1),
\eeq
where $v_{\rm rel}$ is the relative velocity between the two $\text{DM}'$ particles, $\alpha_s$ the QCD coupling constant, and $M_{\rm DM}$ the DM mass (taken to be similar to the mass of $\text{DM}'$). The correct thermal relic abundance is for instance obtained for $M_{\text{DM}'}-M_{\text{DM}}\lesssim 35$\,GeV with $M_{\rm DM}\lesssim \text{few}$\,TeV ($10\,\text{TeV}$) for $\text{DM}'$ that is a color triplet (color octet). 

Since $\text{DM}'$ is colored, this is good news for their production at hadron colliders; the inverse processes to the one in \cref{eq:DM':annih}, such as $q\bar q \to \text{DM}'+\text{DM}'$ should have cross sections of QCD size. On the flip side, though, since the mass splitting between $\text{DM}'$ and $\text{DM}$ are small, only a few 10's of GeV, the production of $\text{DM}'$ results in only soft decay products which are hard to see. That is $\text{DM}'$ would typically decay via $\text{DM}'\to \text{DM}+q$ or $\text{DM}'\to \text{DM}+g$ channels, where  $\text{DM}$ takes away most of the energy in the form of its mass, and escapes detection. What is left as an imprint in the detector are the jets that carry a few 10s GeV of energy and momentum, recoiling against missing momentum of similar size, which is quite a challenging signature to search for experimentally.  In the SUSY examples, $\text{DM}'$ could be one of the squarks, $\tilde q$, which would thus decay via a two body channel,  $\tilde q\to q +\text{DM}$. 

\subsection{Effective operators}
No new physics (NP) has been found at the LHC, which could just mean that NP is simply too heavy to be produced in the $pp$ collisions at the LHC. Integrating out the heavy NP states, we can build an effective field theory (EFT) description of processes at the LHC. For instance, an exchange of a heavy NP  resonance in an $s-$channel, such as a heavy vector $Z'$ that couples to quarks with coupling $g_q$, and to DM with coupling $g_{\rm DM}$, leads to a matrix element
\beq
\begin{split}
- i M=g_q \big(\bar q \gamma^\mu q\big) \frac{g_{\mu\nu}}{\hat s - M_{\rm NP}^2}g_{\rm DM}\big(\bar \chi \gamma^\nu \chi\big) \xrightarrow{\hat s \ll M_{\rm NP}^2} - \frac{g_q g_{\rm DM}}{M_{\rm NP}^2}\big(\bar q \gamma^\mu q\big)\big(\bar \chi \gamma_\mu q\big),
\end{split}
\eeq
where we assumed DM to be a Dirac fermion $\chi$. On the right-hand-side we neglected $\hat s$ compared to $M_{\rm NP}^2$, which is a good approximation at low enough energies when resonances cannot be produced on-shell. The scattering is therefore described by a point-like interaction given by the higher dimension operator (in this case, the dimension 6 operator $\big(\bar q \gamma^\mu q\big)\big(\bar \chi \gamma_\mu q\big)$). If a more precise description is needed, one can keep higher order corrections, suppressed by more powers of $\hat s/M_{\rm NP}^2$. 

The above EFT approach is useful since it covers many possibilities. Keeping all the operators of a given dimension, for instance up to dimension 6 or 7, would cover a large set of possible models that can couple DM to the SM fields. The typical signature  of heavy DM produced through such higher-dimension operators is still a mono-jet or a mono-photon (though other options such as mono-$Z$ or mono-Higgs are also possible). The downside of the EFT based approach is that the EFT breaks down for $\sqrt{\hat s}\sim M_{\rm NP}$, while the sensitivity of the LHC searches for heavy DM is such that the mass scale $M_{\rm NP}$ probed is not very far above the collision energy, $\sqrt{\hat s}$. 

\subsubsection{Simplified models}
\begin{figure}[!t]
\begin{center}
\includegraphics[width=1.0\textwidth]{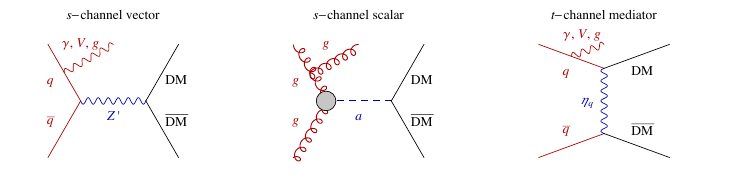}
\end{center}
\caption[DM production via mediator]{\em \label{fig:prod:mediator} 
Representative diagrams for the DM production via $s$-channel (axial-)vector mediator (left panel), $s$-channel (pseudo-)scalar mediator (middle), 
and $t$-channel mediator (right). Reproduced from \cite{Cirelli:2024ssz}.
}
\end{figure}

To deal with this problem one can ``fix'' the EFT by including the relevant degrees of freedom. For DM searches this meant that the description of the DM production at the LHC is amended by including the mediator. The field content at low energies is thus the SM + DM + the mediator. Examples of DM production in commonly introduced simplified models are given in \cref{fig:prod:mediator}.

\begin{table}[t]
\begin{center}
\resizebox{\columnwidth}{!}{
\begin{tabular}{lcccccc}
\hline\hline
\multicolumn{2}{c}{\bf Interaction}   & \multicolumn{1}{c}{\bf Direct detection}  & \multicolumn{1}{c}{\bf Indirect detection}  & \bf Colliders
\\
Type  & ${\cal L}_{\rm int}$ &  NR limit, $\chi A \to \chi A$   & $\chi \bar \chi \to q \bar q$ & $\sigma(q\bar q \to \chi \bar \chi)$
\\[1mm] \rowcolor[rgb]{0.95,0.97,0.97}
\hline
 $V\times V$ & $ \big(g_{\rm DM} \bar \chi \gamma^\mu \chi +g_q \bar q \gamma^\mu q \big) V_\mu$  & $1_\chi 1_N$   & $s$-wave & $\sigma_0 $ 
 \\ \rowcolor[rgb]{0.95,0.97,0.97}
  $V\times A$ & $ \big(g_{\rm DM} \bar \chi \gamma^\mu \chi +g_q \bar q \gamma^\mu \gamma_5 q \big) V_\mu$  & $1_\chi (\vec S_N \cdot \vec v_\perp), \vec S_\chi \cdot (\vec q\times \vec S_N)/m_N$   & $s$-wave & $\sigma_0$ 
  \\ \rowcolor[rgb]{0.95,0.97,0.97}
  $A\times V$ & $ \big(g_{\rm DM} \bar \chi \gamma^\mu \gamma_5\chi +g_q \bar q \gamma^\mu q \big) V_\mu$  & $(\vec S_\chi \cdot \vec v_\perp) 1_N, \vec S_\chi \cdot (\vec q\times \vec S_N)/m_N$   & $p$-wave & $\sigma_0$ 
  \\ \rowcolor[rgb]{0.95,0.97,0.97}
    $A\times A$ & $ \big(g_{\rm DM} \bar \chi \gamma^\mu \gamma_5\chi +g_q \bar q \gamma^\mu \gamma_5 q \big) V_\mu$  & $\vec S_\chi \cdot \vec S_N, (\vec S_\chi \cdot \vec q)(\vec S_N\cdot \vec q)/m_\pi^2$   & $p$-wave & $\sigma_0$ 
  \\  
  \hline
  \rowcolor[rgb]{0.97,0.97,0.93}
   $S\times S$ & $ \big(g_{\rm DM} \bar \chi \chi +g_q y_q \bar q  q \big) a$  & $1_\chi 1_N$   & $p$-wave & $3 y_q^2 \sigma_0/4$
 \\ \rowcolor[rgb]{0.97,0.97,0.93}
  $S\times P$ & $ \big(g_{\rm DM} \bar \chi \chi +g_q y_q \bar q i \gamma_5 q \big) a$  &   $1_\chi (\vec S_N \cdot \vec q)  m_N/m_\pi^2$ & $p$-wave & $3 y_q^2 \sigma_0/4$
  \\ \rowcolor[rgb]{0.97,0.97,0.93}
  $P\times S$ & $ \big(g_{\rm DM} \bar \chi i \gamma_5 \chi +g_q y_q \bar q  q \big) a$  &  $(\vec S_\chi \cdot \vec q) 1_N /M_\text{DM}$   & $s$-wave & $3 y_q^2 \sigma_0/4$
  \\ \rowcolor[rgb]{0.97,0.97,0.93}
    $P\times P$ & $ \big(g_{\rm DM} \bar \chi i \gamma_5\chi +g_q y_q \bar q i \gamma_5 q \big) a$  & $ (\vec S_\chi \cdot \vec q)(\vec S_N\cdot \vec q)m_N/ M_\text{DM} m_\pi^2$  & $s$-wave &$3 y_q^2 \sigma_0/4$
    \\
    \hline \hline
\end{tabular}
}
\end{center}
\vspace{-1ex}
\caption[Comparisons of DM production and annihilation cross sections]{\em\label{tab:different:DMint} {\bfseries Comparison of different heavy DM probes} assuming 
as mediators a vector $V_\mu$ or scalar $a$,
and as DM a Dirac fermion $\chi$ with mass $M$.
The first two columns describe the possible chiral structures of couplings to DM and SM quarks $q$, where
$y_q$ is the SM Yukawa.
The third column lists the leading non-relativistic operators giving rise to DM scattering on nuclei. 
The fourth column indicates whether the $\chi\bar \chi \to q\bar q$ annihilation proceeds through $s$-wave or $p$-wave. 
The last column gives the total cross section for $q\bar q \to \bar \chi \chi$ in the limit where the center of mass energy, $\sqrt{\hat s}$, is much larger than the DM mass, but still much smaller than the mediator mass. 
The partonic cross section $\sigma_0=g_{\rm DM}^2 g_q^2 (\hat s/m_{\rm med}^2)^2 /(6\pi \hat s)$ needs to be convoluted with the parton distribution functions in order to obtain the $pp\to \chi \chi$ cross section, relevant for the LHC. Adapted from \cite{Cirelli:2024ssz}.
}
\end{table}

Since the mediator is now part of the theory, there are many more constraints that become relevant. This includes searches that in principle have nothing to do with DM.  For instance, in the $Z'$ example, \cref{fig:prod:mediator} (left), the new vector resonance can be produced on shell  in $q\bar q\to Z'\to q\bar q$ process, which would appear as a two jet resonance in $pp\to 2j$ collisions. The cross section for $Z'$ production is proportional to $\sigma \propto g_q^2$, and has nothing to do with DM. The couplings to DM only enter as part of the branching ratio for $Z'$ to decay to two-jets instead of two DM particles. If $Z'$ couples only weakly to DM, then $Br(Z'\to 2j)$ can be even close to 100\%. These types of searches thus often lead to the most stringent constraints. 

\subsection{Comparing collider, direct and indirect searches}
So far we discussed only production of heavy DM in colliders. There are two other important ways to search for heavy DM: searches for heavy DM particles scattering in detectors deep underground (``direct detection''), and searches for DM annihilating in different parts of our galaxy, resulting in observable signals in the sky (``indirect detection''). Which of the three probes, collider searches, direct detection or indirect detection, are the most effective, depends on how DM couples to visible matter. 

A comparison of  rates for the example of Dirac DM, for  various forms of vector and scalar interactions with SM quarks, is given in \cref{tab:different:DMint}. We see that, depending on the form of DM interaction, direct detection can be velocity suppressed (such as for $V\times A$ and $A\times V$ interactions) or not (such as for $V\times V$ and $A\times A$). Similarly, the indirect detection rates can be either unsuppressed ($s$-wave), or velocity suppressed ($p$-wave), though the pattern of suppressions is different than for direct detection. The production of DM in colliders, at least for energies well above the DM mass, on the other hand, does not depend on the chiral structure of the interactions.
From this comparison we learn that were we to discover heavy DM, having several different probes will really help us to understand the nature of interactions between DM and the SM fields. This lesson of complementarity will also carry over to the study of light dark matter candidates such as in the models of light dark sectors, including axions and ALPs, that we turn our attention to next.

\section{Models with dark forces/light dark sectors}
\label{sec:models:dark:forces}

\subsection{Dark sector portals} 
Very different collider phenomenology is obtained, if dark matter is much lighter than the electroweak scale. The organizing principle here is minimality - one assumes a minimal field content for the mediators that couple visible and dark sectors. This leads to a limited number of so called {\em portals}. 
That is, we assume that the interactions between the SM and the dark sector are given by 
\beq
\label{eq:portal}
{\cal L}_{\text{portal}}=\sum_i c_i {\cal O}_{i\rm SM}  {\cal O}_{i\rm DS},
\eeq
where ${\cal O}_{i {\rm SM}}$ and ${\cal O}_{i{\rm DS}}$ are local operators in the SM and dark sectors, respectively, and the sum runs over different types of opertors.
On general grounds, we expect that the most relevant portal interactions would be those that are of smallest dimensionality, and are thus the least suppressed by powers of a high-energy scale. The portals for different spins and parity of the mediators are listed in \cref{tab:portals}. The interactions are renormalizable if the mediator is  either spin 0 scalar ({\em ``dark Higgs''} portal), a spin 1 vector ({\em ``dark photon''} or a {\em ``light $Z'$''} portal), or a spin 1/2 fermion  ({\em ``heavy neutral lepton''} portal), where in each case the mediator is not charged under the SM gauge group. If the mediator is 
a spin 0 pseudo-scalar ({\em ``axion''} or {\em ``axion-like pseudo-scalar''}), the interactions are of dimension 5, since the $a$ is assumed to be a pseudo-Nambu-Goldstone boson. The interactions in the axion portal are thus suppressed by a high-energy scale represented by the axion decay constant, $f_a$.

\begin{table}[t]
\begin{center}
\begin{tabular}{lc}\normalsize
\\
\hline\hline
Portal & Interactions
\\[1mm]
\hline
Dark Photon, $A'_\mu \qquad$&  $-\epsilon F_{\mu\nu}'B^{\mu\nu}$
\\
Dark Higgs, $S \qquad$&  $(\mu S+\lambda S^2) H^\dagger H$
\\
Heavy Neutral Lepton, $N \qquad$&  $y_N L H N$
\\
Axion-like pseudo scalar, $a$ & $a F\tilde F/f_a$,  $a G\tilde G/f_a$, $ \big(\bar \psi \gamma^\mu \gamma_5 \psi\big) \partial_\mu a/f_a$
\\
\hline\hline
\end{tabular}
\end{center}
\vspace{-5mm}
\caption[Portals]{\em\label{tab:portals} The lowest dimension portals to the dark sector, 
and sketches of the corresponding interactions. Adapted from \cite{Cirelli:2024ssz}.}
\end{table}

Each of these, $A_\mu'$, $S$, $N$ and $a$, can be the DM, as long as the couplings to the SM are small enough so that these DM candidates are stable on cosmological time scales.  
In most analyses of dark portals, however, the above dark states are assumed to be merely mediators between the dark and visible sectors. 
The dark matter particle is thus usually assumed to be a separate, additional state in the dark sector. A UV complete example of a such DM model is the case where DM is charged 
under a dark U(1) gauge symmetry. The associated dark gauge boson $A_\mu'$ can be much lighter than the DM itself; it can even be massless and result in a long range dark interaction. This is a simple realization of a dark photon portal, with DM a separate state in the dark sector, not the dark photon itself. 

Below we will focus on the axion/ALP portal, and work out the collider phenomenology for it. The important feature of ALP portals is that there are many possible interactions with the SM at dimension 5, so that the phenomenology of ALP portals can be quite rich. Nevertheless, it is important to remember that the axion portal does not cover all possible signatures. For instance, in more general models, DM particle could also be accompanied by a whole set of dark sector states, some of them possibly lighter than the DM itself. Such dark sectors can then lead to an even more diverse set of intricate experimental signatures, than the ones we discuss below.

\subsection{QCD axion vs. ALP at colliders}
The motivation for considering ALPs in more detail is that it is assumed to be a pseudo-Nambu-Goldstone-Boson (pNGB) of a spontaneously broken global $U(1)$ symmetry, so that its mass can be naturally much lighter than the UV scale of symmetry breaking. Small mass has an important practical consequence: ALPs can be produced in currently running and planned experiments, providing us  with a window to high scale physics.  

In our discussion we try to be agnostic about the exact nature of UV physics and allow for a large set of ALP couplings. In this ``bottom-up'' view the most general ALP Lagrangian at the electroweak scale $\mu\simeq \mu_{\rm EW}$ is given by
\begin{equation}
\mathcal{L}_{\text{ALP}}=\frac{(\partial_\mu a)^2}{2}-\frac{m_a^2a^2}{2}+\mathcal{L}_{\text{ALP-gauge}}+\mathcal{L}_{\text{ALP-F}} \label{L_ALP}\, ,
\end{equation}
where we include the most general set of couplings to  the SM fields, $\mathcal{L}_{\text{ALP-gauge}}$, $\mathcal{L}_{\text{ALP-F}}$, unlike in \cref{eq:L:axion}, where only couplings to gluons were displayed. The only assumption at this stage is that ALP mass, $m_a$, is the only source of the shift symmetry breaking. 

\begin{figure}[!t]
\begin{center}
\begin{minipage}{0.34\textwidth}
\includegraphics[width=1\textwidth]{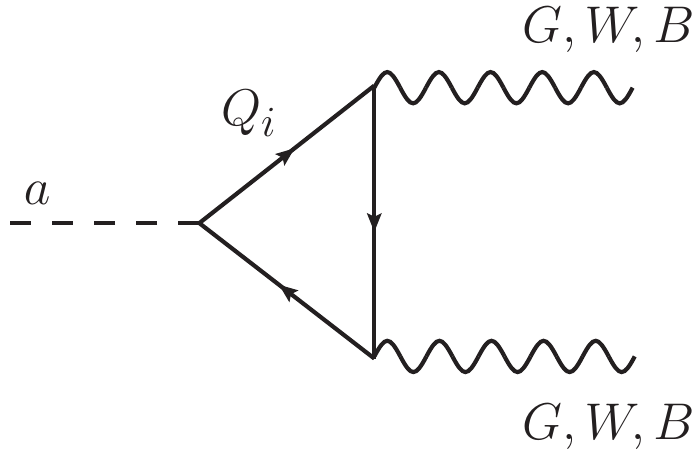}
\end{minipage}
\hspace{0.5cm}
\begin{minipage}{0.1\textwidth}
\huge{$\Rightarrow$}
\end{minipage}
\hspace{0.1cm}
\begin{minipage}{0.26\textwidth}
\includegraphics[width=1\textwidth]{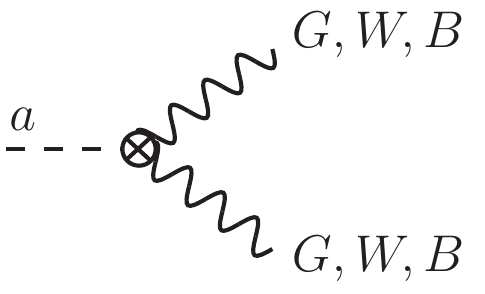}
\end{minipage}
\end{center}
\caption{\em \label{fig:triangle:diagram} 
Triangle diagrams with heavy fermions $Q_i$ or SM fermions running in the loop, give rise to dimension 5 couplings of ALP to gauge bosons,  \cref{eq:alpgauge}, if ALP is a pNGB of an anomalous U(1).
}
\end{figure}

Since ALP is a pNGB, the interactions with the SM fields start at dimension 5.  The couplings to the SM gauge bosons, the gluons, $G^a$, the $SU(2)_L$ gauge bosons, $W^i$, and the hypercharge boson, $B_\mu$, are given by
\begin{equation}\label{eq:alpgauge}
   \mathcal{L}_{\text{ALP-gauge}}=\frac{N_3\alpha_s}{8\pi f_a} a G_{\mu\nu}^a\tilde{G}^{a\mu\nu}+\frac{N_2\alpha_2}{8\pi f_a} a W_{\mu\nu}^i\tilde{W}^{i\mu\nu}+\frac{N_1\alpha_1}{8\pi f_a} a B_{\mu\nu}\tilde{B}^{\mu\nu}\ ,
\end{equation}
where $\tilde G^{a\mu\nu}=1/2\epsilon^{\mu\nu\rho\sigma} G_{\rho\sigma}^a$, with $\epsilon^{0123} = 1$, and similarly for $\tilde W^{i\mu\nu}$, $\tilde B^{\mu\nu}$. These interactions are suppressed by the ALP decay constant $f_a$, whose value roughly coincides with the scale at which the global U(1) symmetry is broken. They are generated from the triangle anomaly diagrams, see \cref{fig:triangle:diagram}, involving the pNGB ($a$) of an anomalous spontaneously broken global $U(1)$ and the SM gauge bosons. In general, they receive contributions from both the SM fermions and from the heavy fermions that were integrated out at $\Lambda_{\text{UV}}$, leading to $N_i\sim {\mathcal O}(1)$.\footnote{For QCD axion, by convention $N_3$ is absorbed into the definition of $f_a$, giving $\mathcal{L}_{\text{QCD axion}}=\frac{\alpha_s}{8\pi f_a} a G_{\mu\nu}^a\tilde{G}^{a\mu\nu}$. We keep the $N_3$ factor explicit.} 

After electroweak symmetry breaking the couplings $W^3$ and $B_\mu$ gauge bosons combine into a massless photon and a massive $Z$, which is integrated out. As a result, at low energies the ALP couplings to gauge bosons only contain couplings to photons
\beq\label{eq:ALPphoton:Lagrangian}
   \mathcal{L}_{\text{ALP-photon}}=\frac{ \alpha}{8\pi f_a} C_{\gamma\gamma} a F_{\mu\nu}\tilde{F}^{\mu\nu},
\eeq 
with $C_{\gamma\gamma}=N_1+N_2$, as well as couplings to gluons, which remain the same as in \cref{eq:alpgauge}.

Since ALP is assumed to be a pNGB, it couples derivatively to the SM fermions,
\begin{equation}\label{eq:ALP-f}
   \mathcal{L}_{\text{ALP-f}} = 
\frac{\partial_\mu a}{2 f_a} \, \bar f_i \gamma^\mu \big( C^V_{f_i f_j} + C^A_{f_i f_j} \gamma_5 \big) f_j  \, , 
\end{equation}
where $f=u,d,e,$ and the sum over repeated generational indices, $i,j=1,2,3$, is understood. The coefficients $C^{V,A}_{f_i f_j}$ are Hermitian matrices in flavor space. In the numerical examples we will take them to be real for simplicity. Because of $SU(2)_L$ gauge invariance we have 
\beq
\label{eq:gauge:rel:CVA}
C^V_{d_i d_j} - C^A_{d_i d_j} = \big(V_{\text{CKM}}\big)^*_{u_k d_i } (C^V_{u_k u_l} - C^A_{u_k u_l}) (V_{\text{CKM}}\big)_{u_l d_j},
\eeq
where $V_{\text{CKM}}$ is a CKM matrix. Furthermore, the $C_{f_i f_i}^V$ have no physical effect. This is straightforward to see by integrating by parts, $\partial_\mu a\bar f\gamma^\mu f=- a \partial_\mu\bar f \gamma^\mu f+\partial_\mu(\ldots)$. Since vector currents are conserved, the first term vanishes, $\partial_\mu\bar f \gamma^\mu f=0$, while the second is a pure derivative and thus contributes only as a vanishingly small boundary term in the action. 

This leads us to the question, how many independent physical parameters are there in the Lagrangian in \cref{eq:ALP-f}? Here, physical means parameters that cannot be rotate away through field redefinitions. Straightforward counting then gives a total of 36 real physical parameters,
\beq
\underbrace{3\cdot (6+3)}_{C_{f_i f_j}^A}+\underbrace{3\cdot (6+0)}_{C_{f_i f_j}^V}-\underbrace{(6+3)}_{\text{CKM rel.}}=36,
\eeq
where a factor of $3$ is for $f=u,d,e$, the first number in each parentheses is for the number of real parameters in the off-diagonal elements, and the second for diagonal elements. Finally, we need to subtract 9 independent linear relations in \cref{eq:gauge:rel:CVA}.

The above ALP Lagrangian is very general, and thus many concrete UV models map onto it. Importantly, this include the QCD axion, which can simultaneously solve the strong CP problem~\cite{Peccei:1977hh,Wilczek:1977pj,Weinberg:1977ma} and account for the observed dark matter abundance~\cite{Preskill:1982cy,Abbott:1980zj,Dine:1982ah}. The QCD axion's mass is only generated from its coupling to the QCD anomaly, is thus small, and given by \cref{eq:QCD:axion:mass} (using the convention where one sets $N_3=1$, otherwise $f_a\to f_a/N_3$ in \cref{eq:QCD:axion:mass}). In general, however, the mass $m_a$ can be viewed as a free parameter. Examples of such ALPs include non-minimal solutions to the strong CP problem, which lead to heavier axions than the naive expectations~\cite{Dimopoulos:1979pp,Holdom:1982ex,Holdom:1985vx,Dine:1986bg,Flynn:1987rs,Choi:1988sy,Rubakov:1997vp,Choi:1998ep,Berezhiani:2000gh,Choi:2003wr,Hook:2014cda,Fukuda:2015ana,Dimopoulos:2016lvn,Agrawal:2017ksf,Agrawal:2017eqm,Gaillard:2018xgk,Hook:2019qoh,Gherghetta:2020keg,Kitano:2021fdl,Gupta:2020vxb}. Other theoretically motivated scenarios  that contain light ALPs include low scale supersymmetry, composite Higgs models, models of dark matter freeze-out, models of electroweak baryogenesis, etc (see, e.g.,~\cite{Kilic:2009mi,Bellazzini:2012vz,Ferretti:2013kya,CidVidal:2018blh,Jeong:2018ucz}). Note, however, that the Lagrangian in \cref{eq:ALP-f} is not the most general interaction Lagrangian for a light scalar - the above Lagrangian assumes $a$ is a pNGB. If this is not the case, and the light scalar couples without derivatives to the SM fermions, then flavor diagonal couplings to scalar currents are also allowed \cite{Delaunay:2025lhl,Batell:2017kty,Batell:2018fqo,Balkin:2024qtf}. 

More often than not the couplings of the QCD axion or the ALPs are assumed to be flavor universal. However, this is not the case in general, and axion/ALP can also have flavor violating couplings. Such flavor violating couplings can simply be a result of the global U(1), that gets spontaneously broken, distinguishing between different generations of the SM fermions. The non-universal U(1) charges can also be due to a deeper reason, as the U(1) can be part of the symmetry that gives rise to the observed hierarchical structure of the SM fermion masses and mixings~\cite{Wilczek:1982rv,Calibbi:2016hwq,Ema:2016ops} (for earlier discussions of flavor-violating axions see Refs.~\cite{Bardeen:1977bd,Davidson:1981zd,Reiss:1982sq,Davidson:1983fy, Davidson:1984ik,Berezhiani:1990wn,Berezhiani:1990jj,Babu:1992cu,Feng:1997tn,Albrecht:2010xh,Ahn:2014gva,Celis:2014iua}). 

The two questions that are phenomenologically very important are tied to the above discussion, and are
\begin{itemize}
\item
Are there sizable flavor violating couplings in the UV, $C_{f_i f_j}^V\sim C_{f_i f_j}^A\sim {\mathcal O}(1)$ for $i\ne j$, or are these instead only generated from the SM CKM off-diagonal matrix elements via loops?
\item
Is the light pseudo-scalar $a$ a QCD axion, and thus invisible in colliders, or is it an ALP, and thus potentially much heavier, and that can decay inside the detectors?
\end{itemize}

The two discrete choices as answers to the above questions lead to drastically different phenomenologies. This is best illustrated on a concrete example, the searches for a QCD axion and ALPs in kaon decays.

\subsubsection{Example: ALP production in kaon decays}

\begin{figure}[!t]
\begin{center}
\begin{minipage}{0.17\textwidth}
\includegraphics[width=1\textwidth]{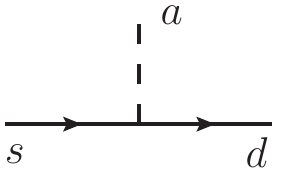}
\end{minipage}
\hspace{1.5cm}
\begin{minipage}{0.31\textwidth}
\includegraphics[width=1\textwidth]{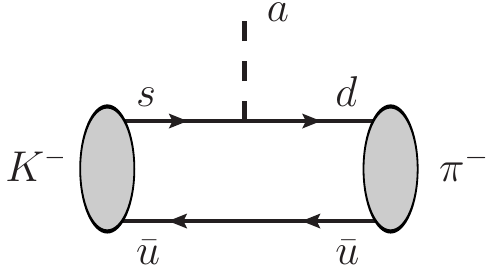}
\end{minipage}
\end{center}
\caption{\em \label{fig:kaon:ALP} 
Partonic $s\to da$ transition (left) induces a $K^-\to \pi^- a$ decay (right).
}
\end{figure} 

\begin{figure}[!t]
\begin{center}
\includegraphics[width=0.28\textwidth]{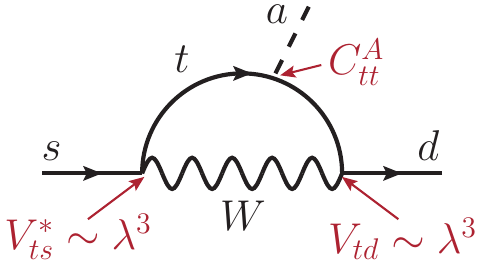}\hspace{0.9cm}
\includegraphics[width=0.28\textwidth]{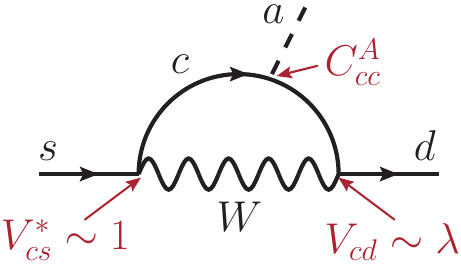}\hspace{0.9cm}
\includegraphics[width=0.28\textwidth]{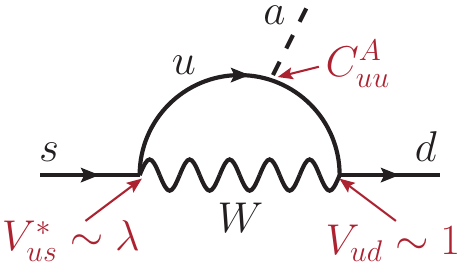}
\end{center}
\caption{\em \label{fig:kaon:loop:gen} 
At one loop the $s\to da$ transition is generated from diagonal couplings of $a$ to up quarks and the $W$ exchange, giving the MFV structure due to CKM  matrix elements.
}
\end{figure}

\begin{figure}[!t]
\begin{center}
\includegraphics[width=0.28\textwidth]{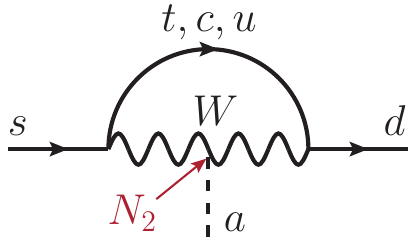}
\end{center}
\caption{\em \label{fig:kaon:loop:gen:W} 
The one loop generated MFV $s\to da$ amplitude due to ALP coupling to $W$ bosons. 
}
\end{figure}

The ALP production in kaon decays is due to a partonic level transition $s\to d a$, which induces the $K^-\to \pi^- a$ decay, where valence quark structure of the two mesons is $K^-\sim [s\bar u]$, and $\pi^-\sim [d\bar u]$, see \cref{fig:kaon:ALP}. The relevant coupling in the ALP Lagrangian in \cref{eq:ALP-f} is $C_{ds}^{V}$, for which we can distinguish between two limiting cases:
\begin{itemize}
\item 
{\bf anarchic} flavor structure: in the UV all couplings of $a$ to the SM fermions are assumed to be large, and thus also $C_{ds}^V(\Lambda)\sim {\mathcal O}(1)$~\cite{Ema:2016ops,Calibbi:2016hwq}.
\item 
{\bf minimal flavor violation (MFV)}: $C_{ds}^V$ is generated from the loops, from either couplings of $a$ to up quarks~\cite{Hall:1981bc,Freytsis:2009ct,Bauer:2020jbp}, see \cref{fig:kaon:loop:gen}, or from couplings of $a$ to $W$ bosons~\cite{Izaguirre:2016dfi}, see \cref{fig:kaon:loop:gen:W}. 
\end{itemize}

At low scale, $\mu\simeq \mu_c\simeq 2\,$GeV,  
the value of $C_{ds}^V$ is therefore given by  \cite{Goudzovski:2022vbt} (shortening from now on $V_\text{CKM}\to V$)
\begin{equation}
\label{eq:CdsV}
C_{ds}^V(\mu_c)=C_{ds}^V(\Lambda_{\text{UV}}) + \sum_{F=t,c}\frac{y_F^2V_{Fd}^*V_{Fs}C_{FF}^{A}}{16\pi^2}\left(\log\frac{\Lambda_{\text{UV}}}{\mu_F}-f_F(x_F)\right)-\frac{g_2^4N_2}{256\pi^4}V_{td}^*V_{ts}f_W(x_t),
\end{equation} 
where $\mu_t=m_Z$, with $x_F=m_F^2/m_W^2$, where $F=t,c$, and the loop functions 
\beq
f_F(x) = \frac{3\left(1-x+\log x\right)}{2(1-x)^2}+\frac{1}{4},
\qquad
f_W(x) = \frac{3 x\left[1-x+x\log x\right]}{2(1-x)^2}.
\eeq
All the Wilson coefficients on the r.h.s. in \cref{eq:CdsV} are defined at the scale $\Lambda_{\rm UV}$, with the second and third terms arising from the loop contributions in \cref{fig:kaon:loop:gen,fig:kaon:loop:gen:W}. The loop contributions are suppressed by the off-diagonal CKM elements, which exhibits a hierarchical pattern 
\beq
V_{\text{CKM}}=
\begin{pmatrix}
V_{ud} & V_{us} & V_{ub}
\\
V_{cd} & V_{cs} & V_{cb}
\\
V_{td} & V_{ts} & V_{tb}
\end{pmatrix}
\sim 
\begin{pmatrix}
1 & \lambda & \lambda^3
\\
\lambda & 1 & \lambda^2
\\
\lambda & \lambda^2 & 1
\end{pmatrix},
\eeq
where $\lambda\simeq 0.2$. This hierarchical structure of the CKM matrix, along with the $1/16\pi^2$ loop factor then gives the highly suppressed numerical values for the loop induced contributions.
Numerically, the sizes of the different contributions in \cref{eq:CdsV} are given by 
\begin{equation}
\label{eq:EWestimate}
\begin{split}
C_{ds}^V(\mu_c)=&C_{ds}^V(\Lambda_{\text{UV}}) 
-10^{-6}\cdot \biggr[(2+i)C_{tt}^A\!\Big(\log\frac{\Lambda_{\text{UV}}}{\mu_t} +0.02
\Big)\!
\\
&\qquad\qquad+\! 0.08 \, C_{cc}^A\!\Big(\log\frac{\Lambda_{\text{UV}}}{\mu_c}+11\Big)\!-(4+2i)10^{-3} N_2\biggr]. 
\end{split}
\end{equation}

The $K\to \pi a$ decay nicely illustrates the stark difference between anarchic and MFV flavor structures. 
If the UV ALP couplings are anarchic, then 
\beq
\label{eq:flavor:anarchy}
C_{ds}^V(\Lambda_{\text{UV}})\sim C_{tt}^V(\Lambda_{\text{UV}})\sim C_{cc}^V(\Lambda_{\text{UV}})\sim N_2.
\eeq
In this case the value of the low scale Wilson coefficients $C_{ds}^V(\mu_c)$, which controls the rate of ALP production in $K\to \pi a$ decays, gets determined almost entirely by the value of the off diagonal coupling in the UV, $C_{ds}^V(\Lambda_{\text{UV}})$. In contrast, the contributions from the diagonal couplings are numerically irrelevant, since they are both loop and CKM suppressed. As we will see below, for $C_{ds}^V(\Lambda_{\text{UV}})\sim {\mathcal O}(1)$ the bounds from $K\to \pi a$ decays translate to a very high scale reach on $f_a$, of order $10^{12}\,$GeV. 

The situation is very different in the case of the MFV flavor structure. Assuming that the flavor violating coupling vanishes in the UV
\beq
C_{ds}^V(\Lambda_{\text{UV}})=0,
\eeq
while the diagonal couplings are of similar sizes 
\beq
 C_{tt}^V(\Lambda_{\text{UV}})\sim C_{cc}^V(\Lambda_{\text{UV}})\sim N_2,
\eeq
we see from \cref{eq:EWestimate} that the low scale coupling $C_{ds}^V(\mu_c)$ is dominated by the top contribution. Furthermore, due to loop and CKM suppression $C_{ds}^V(\mu_c)\ll C_{tt}^V(\Lambda_{\text{UV}})$. As we will see below, for $C_{tt}^V(\Lambda_{\text{UV}})\sim {\mathcal O}(1)$ this translates to bounds on $f_a$ that are in the $f_a\sim 10^6$\,GeV regime, see \cref{fig:BRKpiinv}. 

\subsection{ALP decays}
\label{sec:ALPdecays}
ALP that can be produced in kaon decays need to be fairly light, with a mass less than $m_K-m_\pi\simeq 350\,$MeV.  In this mass range there are only a few kinematically accessible two-body channels; $a\to \gamma\gamma$, the leptonic decays $a\to e^+e^-$, and for $m_a>2m_\mu$ also $a\to \mu^+\mu^-$, and at the very high end of the mass range also the hadronic decay $a\to 2\pi$. For heavier ALPs, produced for instance in $B$ meson or $D$ meson decays, many other decay modes are also possible. 

\begin{figure}[!t]
\begin{center}
\begin{minipage}{0.16\textwidth}
\includegraphics[width=1\textwidth]{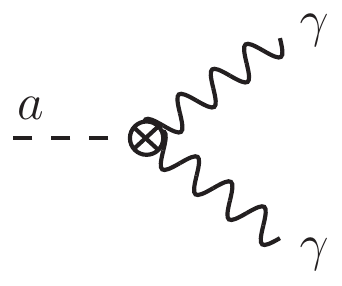}
\end{minipage}
\begin{minipage}{0.27\textwidth}
\includegraphics[width=1\textwidth]{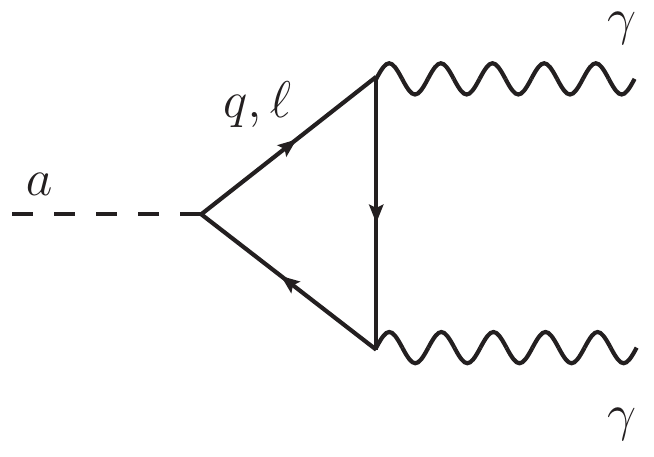}
\end{minipage}
\begin{minipage}{0.33\textwidth}
\includegraphics[width=1\textwidth]{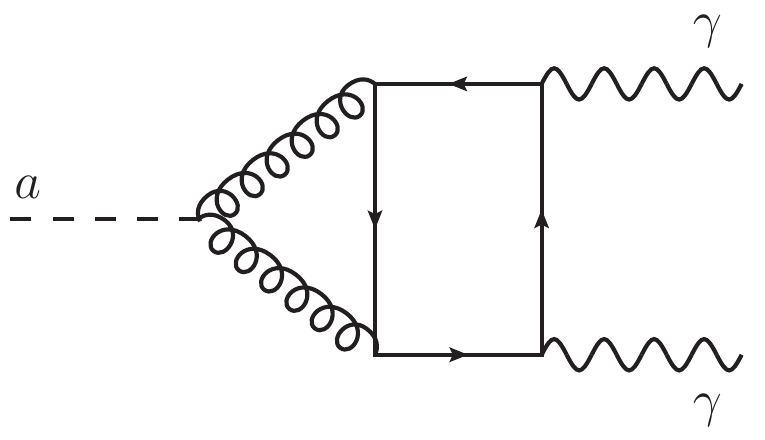}
\end{minipage}
\begin{minipage}{0.21\textwidth}
\includegraphics[width=1\textwidth]{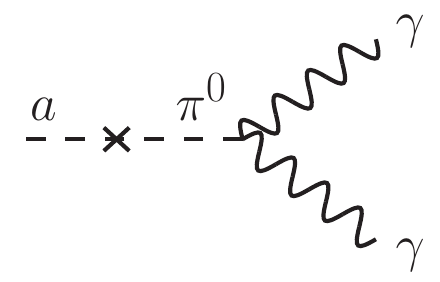}
\end{minipage}
\end{center}
\caption{\em \label{fig:triangle:diagram:plus} 
Different contributions to  $C_{\gamma\gamma}^\text{eff}$ in \cref{eq:cgamgameff}, from the UV couplings; the SM quarks and leptons running in the loop; the nonperturbative contribution from coupling to gluons; and from $a-\pi^0$ mixing. 
}
\end{figure}

The most striking signatures come from decays to leptons and to two photons.  For very light ALP, with a mass below the $2 m_e$ threshold, the only allowed decay is to two photons, with the decay width
\beq\label{eq:ALP-widths}
\Gamma(a\to \gamma\gamma)
= \frac{\alpha^2\, m_a^3}{256\pi^3 f_a^2} |C_{\gamma\gamma}^\text{eff}|^2 \,,
\eeq
where the effective coupling of the ALP to photons has the same normalization as in \cref{eq:ALPphoton:Lagrangian}. (A common notation in axion literature is also $g_{\gamma\gamma}=C_{\gamma\gamma}^\text{eff} \alpha /(2\pi f_a)$.) The ALP-photon coupling receives several different contributions \cite{Bauer:2017ris,Bauer:2020jbp,GrillidiCortona:2015jxo}
\beq
\begin{split}
\label{eq:cgamgameff}
   C_{\gamma\gamma}^\text{eff}
   &= N_1+N_2 +\sum_{q}\, 6\, Q_q^2\,C_{qq}^A(\mu_0)\,B_1(\tau_q)
    + 2\sum_{\ell} C_{\ell\ell}^A\,B_1(\tau_\ell)
    \\
   &\quad\mbox{}- (1.92\pm 0.04)\,N_{3} - \frac{m_a^2}{(m_\pi^2-m_a^2)} 
    \left[ N_{3}\,\frac{m_d-m_u}{m_d+m_u} + C_{uu}^A-C_{dd}^A \right] \,.
    \end{split}
\eeq

The first two terms, $N_1+N_2$, are just the linear combination of UV couplings to $B_{\mu\nu}$ and $W_{\mu\nu}^a$ that give the electromagnetic $F_{\mu\nu}F^{\mu\nu}$ combination after the electroweak symmetry breaking (see left diagram in \cref{fig:triangle:diagram:plus}). The next two terms are from heavier quarks running in the loop, and from the leptons running in the loop, respectively, see the second diagram in \cref{fig:triangle:diagram:plus}.  
The loop function $B_1(\tau_f)$ depends on $\tau_f\equiv 4m_f^2/m_a^2$, and is $B_1\simeq 1$ for light fermions ($m_f\ll m_a$) while it decouples as $B_1\simeq -m_a^2/(12 m_f^2)$ for heavy fermions ($m_f\gg m_a$). The explicit form of $B_1(\tau)$ can be found in Ref.~\cite{Bauer:2020jbp}. 
The second line in Eq.~(\ref{eq:cgamgameff}) encodes the nonperturbative contributions from couplings to gluons and light quarks. These can be estimated in ChPT, are of two types. The first is the point-like contribution from $a$ coupling to gluons, which in the perturbative picture would be a two-loop effect, from the gluon loop attaching to the quark loop, which then can emit photons, see third diagram in \cref{fig:triangle:diagram:plus}. 
In \cref{eq:cgamgameff} we show the numerical value for the NLO ChPT result from \cite{GrillidiCortona:2015jxo}. The last term in \Cref{eq:cgamgameff}, is from $\pi^0-a$ mixing, with $\pi^0$ coupling to photons via WZW anomaly, see the last diagram in \cref{fig:triangle:diagram:plus}.  

Numerically, the decay length for the case where $a\to \gamma\gamma$ decays dominate, turns out to be very long
\begin{align}
    c\tau_a 
    \approx 2.9\times10^7\,{\rm meters} \left( \frac{f_a}{\rm TeV} \right)^2 
    \left(\frac{1\,{\rm MeV}}{m_a} \right)^3
    \left(\frac{1}{|C_{\gamma\gamma}^\text{eff}|}\right)^2  \,,
    \label{eq:ALPlifetime:photon}
\end{align}
where as an example we used a relatively low scale for $f_a$, and ALP mass $m_a$ at the $2m_e$ threshold, i.e., the heaviest ALP mass before $a\to e^+e^-$ decays become kinematically allowed. For larger $f_a$ and for lighter ALP the $c\tau_a$ are even longer.  This means that an ALP lighter than the di-electron threshold will not decay within the detector, unless $f_a$ is extremely small.

For heavier ALP, the decay channel to lepton pairs opens up.  The corresponding decay width is
\beq
\Gamma(a\to\ell^+\ell^-) 
=   \frac{m_a\, m_\ell^2}{8\pi f_a^2}\,\big(C_{\ell\ell}^{A} \big)^2\,\Big(1-\tfrac{4m_\ell^2}{m_a^2}\Big)^{1/2} \,.
\eeq
For ${\mathcal O}(1)$ ALP couplings to electrons the corresponding ALP decay lengths differ drastically, whether or not the decays to heavier leptons are kinematically allowed, due to $m_\ell^2$ scaling,
\begin{equation}
\label{eq:ctau_a:leptons}
 c\tau_a 
    \approx
    \left\{
    \begin{aligned}
        & 1.9\text{ meters}\left( \frac{f_a}{\rm TeV} \right)^2 
    \left(\frac{10\,{\rm MeV}}{m_a} \right)
    \left(\frac{1}{C_{ee}^A}\right)^2\ , & 2 m_e<m_a< 2 m_\mu,
        \\
        & 1.5\times 10^{-10}\text{ meters}\left( \frac{f_a}{\rm TeV} \right)^2 
    \left(\frac{300\,{\rm MeV}}{m_a} \right)
    \left(\frac{1}{C_{\mu\mu}^A}\right)^2\ ,& 2 m_\mu<m_a<2m_\tau.
    \\
         & 4\times 10^{-6}\text{ meters}\left( \frac{f_a}{\rm TeV} \right)^2 
    \left(\frac{4\,{\rm GeV}}{m_a} \right)
    \left(\frac{1}{C_{\tau\tau}^A}\right)^2\ ,& 2 m_\tau<m_a.
    \end{aligned}
    \right.
\end{equation}
The above lifetimes assume that couplings to leptons dominate. Especially for heavier ALPs, such that $a\to 2\tau$ decays are allowed, it is very likely though that other channels, such as $a\to{\rm hadrons}$, are open as well.

The above numerical results show that for $f_a$ and $m_a>2 m_e$ ALPs decay inside the detector, assuming ${\mathcal O}(1)$ dimensionless couplings. For $f_a\gg \text{TeV}$ and/or $m_a\ll 1$\, MeV, on the other hand, ALPs are stable on detector scales and appear as missing energy and momentum in the experiment.

\subsubsection{Example: QCD axion and ALP constraints from kaon decays}
Let us now focus on an explicit example: the constraints that are being placed on QCD axions and ALPs from rare kaon decays. 

We start with the QCD axion case. The QCD axion mass is generated from its couplings to gluons. Since these couplings are highly suppressed, the induced QCD axion mass is also very small, $m_a\ll 2m_e$, \cite{GrillidiCortona:2015jxo} 
\beq
m_a=5.7\,\mu\text{m}\, \Big(\frac{ 10^{12}\,\text{GeV}}{f_a/N_3}\Big).
\eeq
If QCD axion has flavor violating couplings, so that transitions $s\to d a$ are possible,  it will therefore appear as an invisible momentum in the two body kaon decay,
\beq
K^-\to \pi^- a_{\rm inv}.
\eeq
Since the invisible mass is much smaller than the energy resolution of the experiments, one can thus effectively take $m_a\approx 0$. A defining feature of such decays is a mono-energetic $\pi^-$ in kaon rest-frame, with the energy $E_\pi \simeq m_K/2$ (or, equivalently, an invisible mass peak at $(p_K-p_\pi)^2\approx 0$).   
The SM background is due to the very rare three body decays $K^-\to \pi^-\nu\bar \nu$ in which pion can take a range of energies. The kinematics of this background is thus  very different from the monoenergetic signature of the $K^-\to \pi^- a_{\rm inv}$ decays. 

\begin{figure}[t]
\centering
\includegraphics[width=0.49\textwidth]{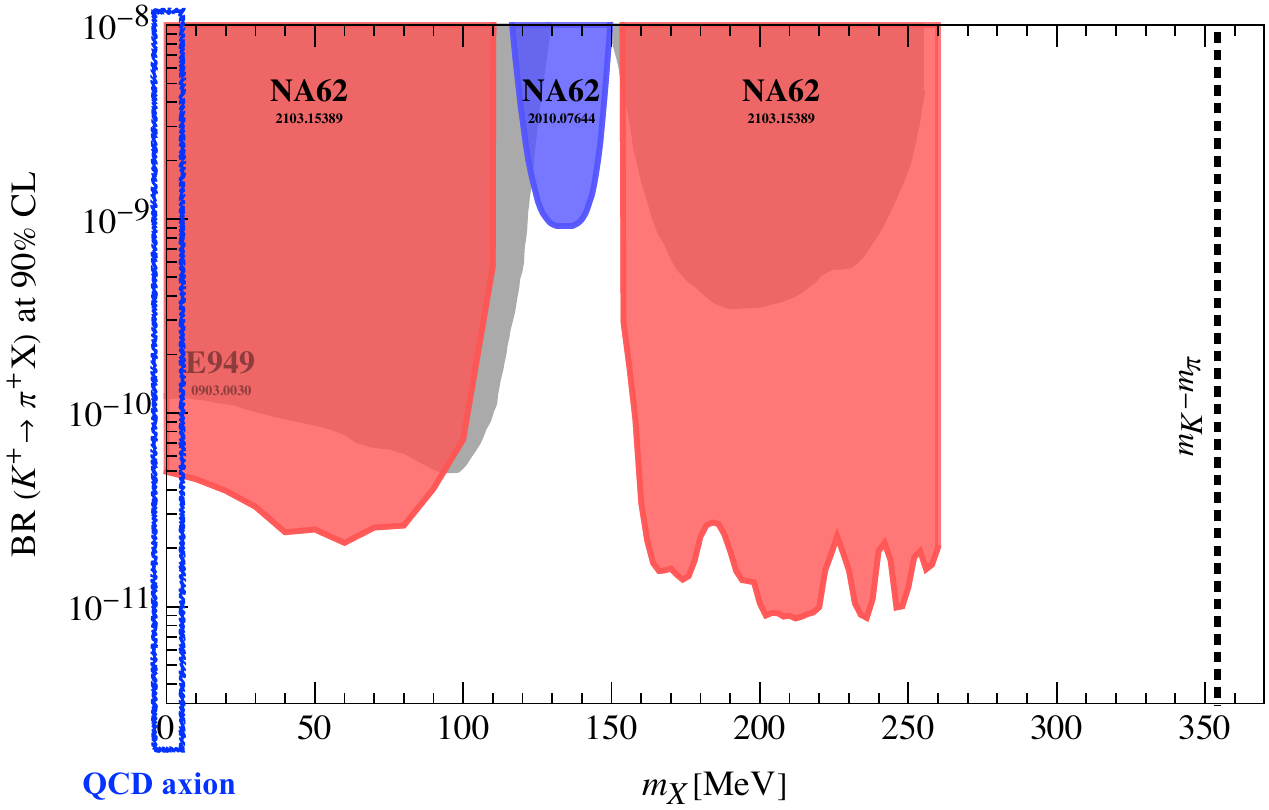}
\includegraphics[width=0.49\textwidth]{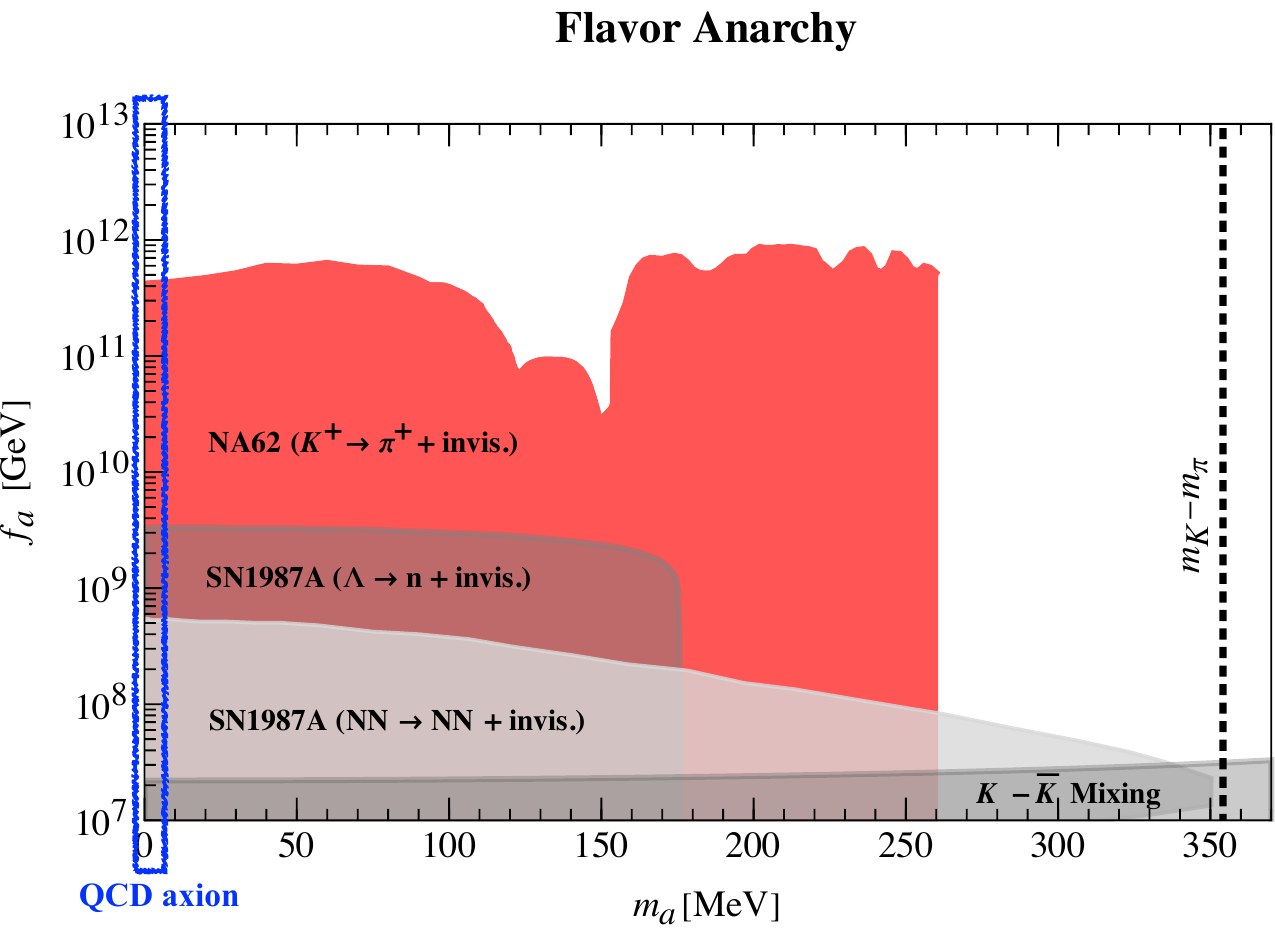}
\caption{\em Left: Constraints on ${\cal B}(K^+\to\pi^+X)$ with the $X$ particle decaying invisibly or escaping the detector as a function of $X$ mass up to the kinematic endpoint (vertical dotted line). Colored regions are excluded, from $K^+\to\pi^+X$ search~\cite{NA62:2021zjw} (red) and~\cite{BNL-E949:2009dza} (grey), and by $\pi^0\to{\rm inv}$ search~\cite{NA62:2020pwi} (blue).  
Right: Constraints on the axion decay constant $f_a$ for the flavor anarchic case, from  $K^+\to\pi^+X$ search~\cite{NA62:2021zjw} (red), and the SN1987A cooling constraints~\cite{Lee:2018lcj,MartinCamalich:2020dfe} (grey). The $K-\bar K$ mixing constraints assume that the dimension-6 UV contributions are small. The blue boxes higlight the region of QCD axion mass. Adapted from \cite{Goudzovski:2022vbt}.
\label{fig:BRKpiinv}}
\end{figure}

For heavier ALP, the two-body $K^-\to \pi^- a$ decays still lead to mono-energetic pions, but now with a smaller energy. \Cref{fig:BRKpiinv} (left) shows bounds on branching ratio $Br(K^+\to \pi^+a)$, assuming that also heavier ALPs remain invisible and do not decay in the detector. In most of the kinematically accessible $m_a$ range, the bound on $Br(K^+\to \pi^+a)$  is significantly below the $10^{-10}$ level. This is to be compared  with the SM background due to the $K^+\to \pi^+\nu\bar\nu$ three body decays, which have a branching ratio 
$Br(K^+\to \pi^+\nu\bar\nu)=(1.14\pm0.36)\cdot 10^{-10}$. 

In order to translate the bounds on $Br(K^+\to \pi^+a_{\rm inv})$ into bounds on $f_a$, one needs to assume a particular flavor structure. For this,  let us make use of our two representative cases. 

\paragraph{Anarchic flavor structure.} Assuming that the couplings satisfy \cref{eq:flavor:anarchy}, where we set all couplings also equal to 1, the bounds on $f_a$ as a function of ALP mass are shown in \cref{fig:BRKpiinv} (right). We see that very high scales get probed. What is the reason for this? The first is that the kaon decay width, dominated by $Br(K^+\to \mu^+\nu_\mu)\simeq 64\%$ and $Br(K^+\to \pi^+\pi^0)\simeq 21\%$, is highly suppressed. Any exotic branching ratio is measured relative to this small SM decay width, induced by the tree level $W^\pm$ exchange. Integrating out the heavy $W$, since $m_W\gg m_K$, gives an effective Lagrangian for the semileptonic decay that is of dimension six
\beq
{\cal L}_{\rm eff}=\frac{g^2}{m_W^2} V_{us}\big(\bar u\gamma^\mu s\big)_{V-A} \big(\bar \mu \gamma_\mu \nu_\mu \big)_{V-A},
\eeq
with $\big(\bar \psi \gamma^\mu\psi'\big)_{V-A}=\big(\bar \psi \gamma^\mu(1-\gamma_5)\psi'\big)$. The analysis for hadronic decays of kaons can also be performed using the appropriate dimension six effective Lagrangian, though the calculation in that case is more involved, since there are two hadrons in the final state.
Parametrically, the SM decay width for the charged kaon is
\beq
\Gamma_K\propto \frac{m_K^5}{m_W^4}\simeq 10^{-9}\cdot m_K,
\eeq
where the undisplayed proportionality factor contains additional suppressions such as due to $(m_\mu/m_K)^2\simeq 0.05$, $V_{us}^2\simeq 0.05$, and the suppressions due to nonperturbative hadronic physics.

\begin{figure}[t]
\centering
\includegraphics[width=3in]{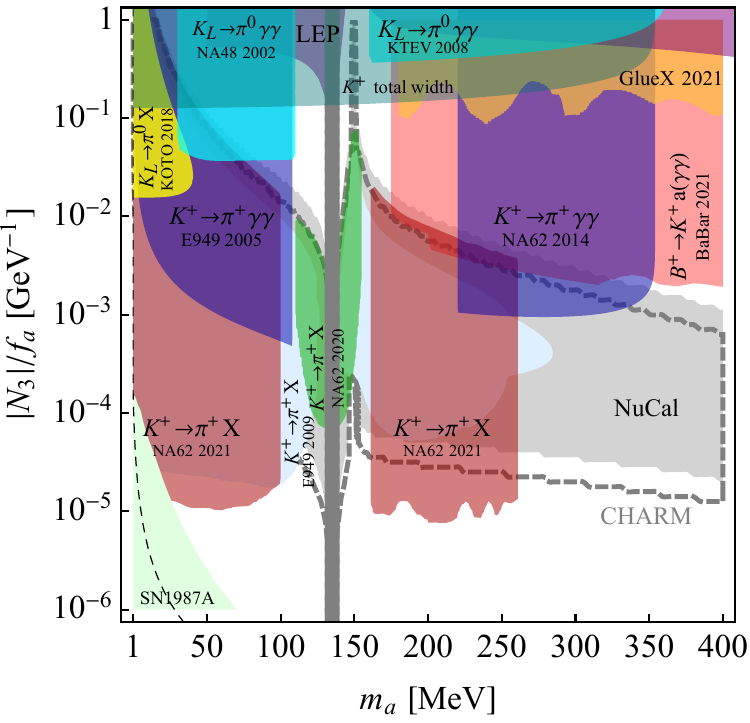}
\caption{\label{fig:cgg:bounds}
{\em  Bounds on the ALP coupling only to gluons in the UV ($N_3(\Lambda_{\rm UV})\ne 0$ only). Reproduced from \cite{Goudzovski:2022vbt}.}}
\end{figure}

\paragraph{MFV flavor structure.} If the off-diagonal couplings are zero in the UV, $ C_{ij}^V(\Lambda_{\text{UV}})=0$ for $i\ne j$, these are generated only through loops and are thus highly suppressed, see \cref{eq:CdsV}. As the result the searches in rare kaon decays now probe much smaller UV scales. An example is shown in \cref{fig:cgg:bounds}, where we assumed that in the UV ALP only couples to gluons. As we see, the bounds from kaon decays are in this case
\beq
\frac{f_a}{N_3}\lesssim 10\,\text{TeV},
\eeq
which is quite representative also of other choices for flavor diagonal couplings in the UV. Since the suppression scale $f_a$ is relatively low, the ALP can now decay inside the detectors. Along with $K^+\to \pi^+ a_{\rm inv}$ and $K_L\to \pi^0 a_\text{inv}$, one can now also search for ALPs in the $K^+\to \pi^+ (a\to \gamma\gamma)$ and $K_L\to \pi^0 (a\to \gamma\gamma)$ channels (the decays $a\to e^+e^-$ are relevant only in a limited part of parameter space).

\subsection{Brief summary of other ALP collider searches}
Let us briefly review other possible searches for ALP in colliders. 

\paragraph{Rare meson decays.} If ALPs are heavy enough, $K^+\to \pi^+a$ decays will be kinematically forbidden. Decays of  mesons composed out of heavier quarks, such as $B^+\to K^+ a$ or $D^+\to \pi^+a$, etc, may then become relevant. As for the case of rare kaon decays, discussed in the previous section, the UV scales probed depend on the assumed flavor pattern of $\bar q_i q_j a$ couplings. For anarchic flavor structure the $B^+\to K^+ a$, $D^+\to \pi^+a$, etc, decays will probe high scales, $f_a$ above $10^{7-8}\,\text{GeV}$ \cite{MartinCamalich:2020dfe}, while for MFV case the probed $f_a$ scale is in the TeV range. For very small masses, $a$ escapes the detector, while heavier $a$ will decay inside the detector. The difference with the rare kaon decays is that there are many more $a$ decay channels that are kinematically open. The branching ratio to the  channels that can be easily searched for, such as $a\to 2e, a\to 2\gamma$ or $a\to 2\mu$ may thus be subleading, if, for instance, $a$ is hadrophilic.

\paragraph{In $e^+e^-$ collisions.} The prototypical example here is the ALP coupling to photons, in which case this can be produced in a process $e^+e^-\to \gamma^*\to a \gamma$ via a diagram shown in \cref{Tutorial:ALPphotonATeecolliders:ProductionDiagrams} (left). 
The ALP can then decay to two photons, $a\to \gamma\gamma$. If $a$ is relatively light, then the two photons would be highly collimated and may not get resolved by the experiments, appearing as an additional photon. A bound on such ALP for instance comes from LEP inclusive $e^+e^-\to 2\gamma$ searches. We will discuss this type of signatures in detail in the tutorial, in \cref{sec:tutorial} below.

A different possibility is that $a$ is so light that it does not decay within the detector. The process $e^+e^-\to \gamma^*\to a \gamma$, with $a$ escaping, then leads to a {\em monophoton} signature (single photon + missing energy).

The ALP can also be searched for in radiative Upsilon decays, where $\Upsilon(nS)$ (a $b\bar b$ bound state) are created in $e^+e^-$ collisions, typically when the energy of collisions is tuned to the resonance mass. The Upsilon then decays via $\Upsilon(nS)\to \gamma^*\to \gamma +a$. If ALP escapes detection, this again leads to the monophoton signature. 

Finally, consider the $a Z_{\mu\nu}\tilde F^{\mu \nu}$ coupling, where $Z_{\mu\nu}=\partial_\mu Z_\nu -\partial_\nu Z_\mu$, with $Z_\mu$ the $Z$-boson gauge field. The $e^+e^-$ collisions on the $Z$ resonance peak at LEP could then lead to the $e^+e^-\to Z\to a\gamma$ process, and thus a monophoton signal at LEP.

\paragraph{In $pp$ collisions.} If ALP couples to both gluons and photons, then it can be produced in $pp$ collisions through its gluon couplings, and be searched via its $a\to \gamma\gamma$ decays \cite{CidVidal:2018blh} (the $a\to {\rm hadrons}$ decays, that would dominate for heavier ALP masses, are very hard to search for in the very busy hadronic environment of $pp$ collisions).

\paragraph{In ion-ion collisions.} An interesting and qualitatively different possibility is production of ALP in peripheral ion-ion collisions, $A+A\to A+A$ \cite{Knapen:2016moh}. In peripheral collisions ions only brush up against each other, and remain intact. Such a collision results in a very clean event, with only two ions getting deflected at an angle. It can also, however, result in a coherently enhanced production of a light ALP through its couplings to photons. Since the momentum exchange is small, the ALP ``sees'' the whole nucleus, leading to 
\beq
\sigma(A+A\to A+A+a)\propto Z^4,
\eeq
where $Z e$ is the charge of the nucleus $A$. For heavy nuclei, such a lead with $Z=82$, this coherence enhancement can be large, ${\mathcal O}(10^8)$. The signal in the detector is the same as for normal peripheral ion-ion collisions, two deflected nuclei, but now in addition with remnants of ALP decays, such as $a\to \gamma\gamma$. 

\paragraph{Proton beam dumps.}  We have seen that $a G\tilde G$ couplings can induce production of ALPs via rare meson decays. Proton beams that collide with a thick target (``proton beam dump'') can have high intensity and thus lead to large samples of various mesons: kaons, $B$ mesons, $D$ mesons, etc... (The production of strange and heavy flavor baryons is suppressed by about an order of magnitude compared to production of mesons, and can thus to first approximation be neglected.)  The ALP can then be produced in rare $K\to \pi a$, $B\to Ka$, ... decays. The thick target is followed by a thick shield, ``'dirt'', which can be even 100s of meters thick, after which one places a detector. The ALPs, if they are long lived enough, will traverse dirt without interacting, and can then decay in the detector (see schematics in \cref{fig:beam:dump}). The dominant decay channel depends on the details of how ALP couples to the SM fermions, see \cref{sec:ALPdecays}. To get rough understanding of the bounds let us consider the simplified case where ALP predominantly couples to gluons, and thus the dominant channels are $a\to \gamma\gamma$ for small masses and $a\to {\rm hadrons}$ for heavy masses. 

\begin{figure}[t]
\centering
\includegraphics[width=0.5\textwidth]{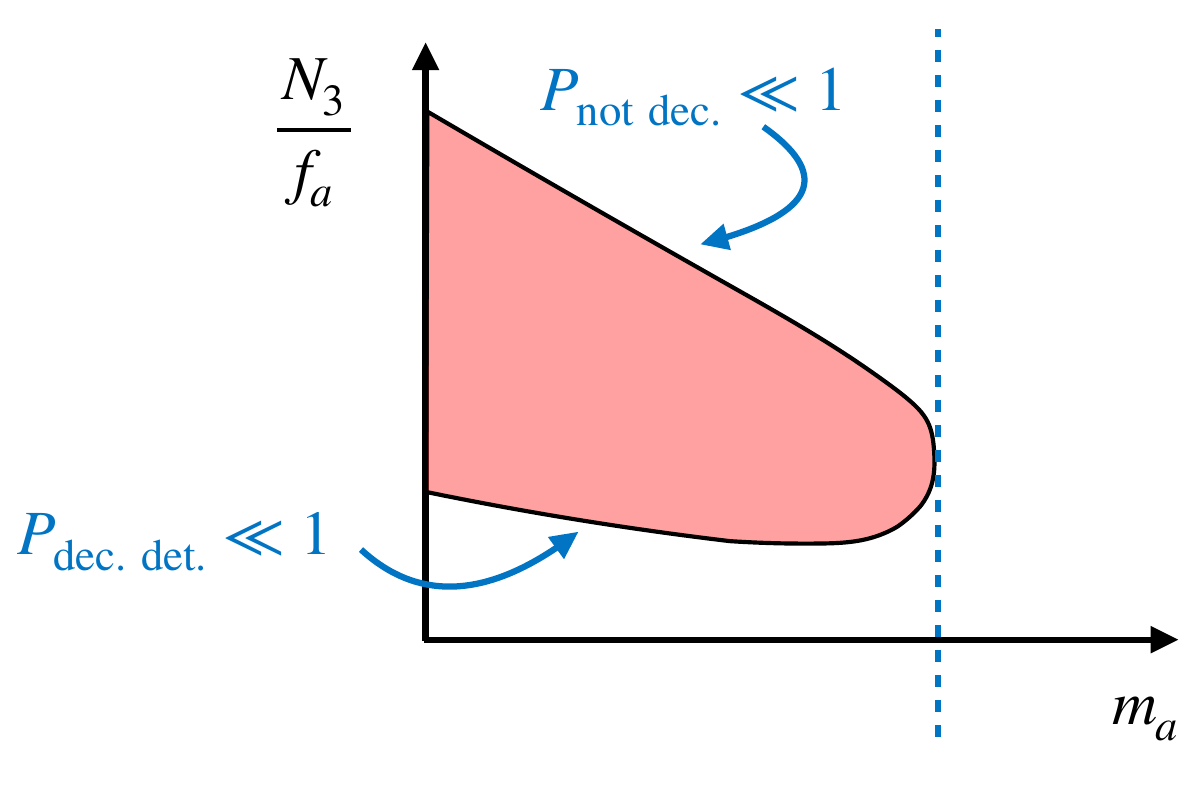}
\caption{\label{fig:scaling:bounds}
{\em  Typical shape of exclusions from beam dump experiments on ALP couplings to gluons $N_3/f_a$ as a function of ALP mass $m_a$. Dashed vertical line gives the kinematical endpoint.}}
\end{figure}

The typical shape of exclusion is shown in \cref{fig:scaling:bounds}. The upper boundary, for small values of $f_a/N_3$, and thus relatively strong couplings of ALP to gluons, is obtained when most of the ALP decay within dirt and do not make it all the way to the detector. The lower boundary of the excluded region, on the other hand, is obtained when the number of ALPs produced in the beam dump becomes too small to give an observable signal, because the coupling $N_3/f_a$ is too small. The highest $m_a$ reach on the other hand is given by how high the energy of the protons is, in particular whether or not $B$ mesons can be produced, or is the ALP production instead due to the decays of lighter mesons, $D$'s or even only $K$'s.

We can be slighly more quantiative, and also obtain the scalings with $f_a$. The number of produced ALPs, $N_{\rm prod}$, is proportional to
\beq
N_\text{prod}\propto \biggr(\frac{N_3}{f_a}\biggr)^2.
\eeq
The decay width, on the other hand, is
\beq
\label{eq:Gamma:a}
\Gamma_a=\frac{1}{c\tau_a}\propto \biggr(\frac{N_3}{f_a}\biggr)^2 m_a^3.
\eeq
The number $N_{\rm obs}$ of observed ALP decays in the detector is given by
\beq
\label{eq:Nobs}
N_\text{obs}=N_\text{prod}\cdot P_\text{not dec.}\cdot P_\text{dec. det.},
\eeq
where the probability of ALP not decaying before reaching the detector reads
\beq
P_\text{not dec.}= \exp\Big(-L_\text{dirt}/c\tau_a \gamma\Big),
\eeq
with $L_\text{dirt}$ the thickness of the dirt shield, and $\gamma$ the typical Lorentz boost factor of the ALPs produced in the beam dump. The remaining factor in \cref{eq:Nobs} gives the probability for the particle to decay inside the detector (after it has reached the detector), and is
\beq
P_\text{dec. det.}= 1- \exp\Big(-L_\text{det}/c\tau_a \gamma\Big)\approx L_\text{det}/c\tau_a \gamma,
\eeq
where in the last approximation we assumed that the linear dimension of the detector, $ L_\text{det}$, is much smaller than the effective decay length, $c\tau_a \gamma$. For ALP couplings to gluons, thus 
\beq
P_\text{dec. det.}\propto L_\text{det} \biggr(\frac{N_3}{f_a}\biggr)^2 m_a^3, 
\eeq
For large $f_a$ and/or small $m_a$ we can quickly reach the limit $ L_\text{det}\ll c\tau_a \gamma$. If also $L_{\rm dirt}\ll c\tau_a \gamma$, so that $P_{\rm not dec.}\approx 1$, then 
\beq
N_{\rm obs}\propto \biggr(\frac{N_3}{f_a}\biggr)^4m_a^3,
\eeq
where two powers of $1/f_a$ suppression come from the reduced production, and two powers from the reduced probability for the particle to decay inside the detector volume. 
Assuming that a discovery of an ALP requires some fixed value of observed events, this explains the lower boundary of the excluded region in \cref{fig:scaling:bounds}, $N_3/f_a\big|_{\rm excl. low.}\propto m_a^{-3/4}$. The upper boundary of the excluded region, on the other hand, is reached when the coupling $N_3/f_a$ is so large that it becomes very unlikely for ALP to even reach the detector, i.e., in the limit when 
\beq
P_\text{not dec.}= \exp\Big(-L_{\rm dirt}\Gamma_a/\gamma\Big)=\exp\Big(-\# (N_3/f_a)^2 m_a^3\Big) \ll 1.
\eeq
This then gives  a rough scaling for the upper exclusion boundary $N_3/f_a\big|_{\rm excl. up.}\propto m_a^{-3/2}$.

\begin{figure}[t]
\centering
\begin{minipage}{0.20\textwidth}
\includegraphics[width=1.\textwidth]{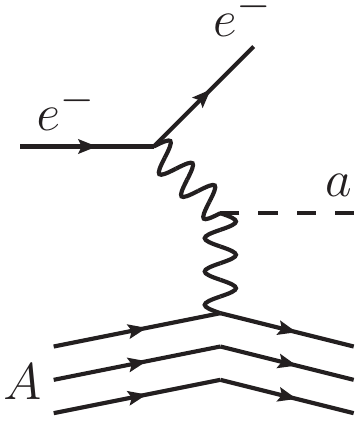}
\end{minipage}
\hspace{2cm}
\begin{minipage}{0.20\textwidth}
\includegraphics[width=1.\textwidth]{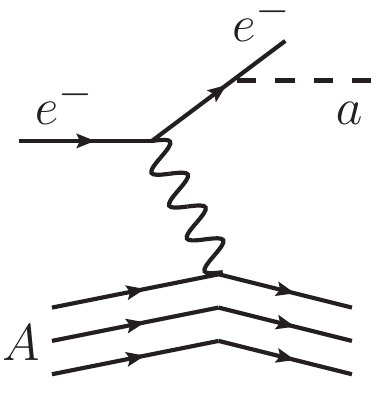}
\end{minipage}
\caption{\label{fig:Primakoff}
{\em  Production of ALPs in electron beam dumps from Primakoff process (left) and from electron bremsstrahlung (right). }}
\end{figure}

\paragraph{Electron beam dumps.} In electron beam dumps the ALPs are produced in a Primakof process, where $e^-$ in the electro-magnetic field of nucleus emits an ALP due to its couplings to photons, see the diagram in \cref{fig:Primakoff} (left). If ALP couples to electrons also a bremsstrahlung emission from electron leg contributes to the flux, \cref{fig:Primakoff} (right).  The experimental set-up is similar to the one for proton beam dumps with a detector placed behind a shielding that stops most of the backgrounds from SM particles that also get produced in target. The dominant ALP channels again depend on the details of ALP couplings. For light ALP though, the dominant decays will be $a\to \gamma\gamma$ and $a\to e^+e-$ simply due to kinematics. 

\section{Tutorial}
\label{sec:tutorial}

Physics is, at its core, an experimental science. Any new theoretical framework thus must ultimately lead to testable predictions -- signals or signatures that can be observed in current or future experiments. In other words, for a theory to be scientifically meaningful, it must be falsifiable.

In this last part of the lecture notes we provide a practical tutorial on how to translate a theoretical model of particle interactions into predictions. 
In the context of particle physics, this typically involves the following steps:
\begin{enumerate}
\item write down the Lagrangian that describes the new physics interactions,
\item calculate the relevant scattering cross sections and decay widths,
\item compute the expected event rates for the experimental setup under consideration,
\item confront theoretical predictions with experimental data.
\end{enumerate}
While the first two steps are standard fare in most particle physics courses,
the last two require some knowledge of the relevant experiments. For instance, it is important to understand what the experiment was optimized for, what it can and cannot measure. For collider experiments, we discussed  this in \cref{sec:BasicsOfCollider}. The endgoal is to deduce which regions of parameter space can be probed experimentally, given the data collected and the systematic errors. 
If statistically significant deviations from the SM predictions are found, one can then try to interpret data within a specific new physics model; conversely, if no excess is found, one can compute exclusions in the space of model parameters.

In the remainder of this section we will illustrate the above procedure on a concrete example: the search for ALP--photon interactions at Belle II. The sequence of steps involved will be quite common for a typical phenomenological analysis, even though this is not intended to be a detailed  study; for that, we refer the reader to Ref.~\cite{Belle-II:2020jti}, which serves as a basis for this tutorial.

\subsection{ALP-photon probes at $e^+e^-$ colliders}
\label{sec:Tutorial:ALPphotonATeecolliders}

\begin{figure}[t]
	\centering
    \includegraphics[width=.6\linewidth]{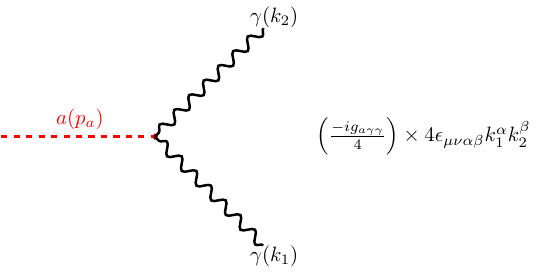}
    \caption{Feynman diagrams for the ALP decay at tree level, with the associated Feynman rule for the vertex. }
\label{Tutorial:ALPphotonATeecolliders:FeynmanRule}
\end{figure}

In order to follow the notation in Ref.~\cite{Belle-II:2020jti} more closely, let us use a slightly different normalization of ALP--photon couplings than in \cref{eq:ALPphoton:Lagrangian}, and define
\beq\label{eq:Tutorial:ALP-photon:Lagrangian}
{\cal L}_{\rm ALP} \supset \frac{g_{a\g\g}}4 aF_{\mu\nu}\tilde F^{\mu\nu}\,,
\eeq
where $a$ is the ALP, $F_{\mu\nu}$ is the photon field strength, and its dual is defined as $\tilde F^{\mu\nu} = \frac12\epsilon^{\mu\nu\alpha\beta}F_{\alpha\beta}$. Note that the coupling $g_{a\g\g}$ has dimensions of $\text{GeV}^{-1}$.
This is equivalent to the Lagrangian given in \cref{eq:ALPphoton:Lagrangian}, in which, however this mass-scale dependence was made explicitly through the appearance of $1/f_a$. 
The first step in our analysis is to derive the Feynman rule associated with the $a\g\g$ vertex. For Lagrangian in \cref{eq:Tutorial:ALP-photon:Lagrangian} the Feynman rule is given by
\beq\label{eq:Tutorial:ALP-photon:FeynRule}
\lp \frac{-ig_{a\g\g}}4 \rp\times4\epsilon_{\mu\nu\alpha\beta}k_1^\alpha k_2^\beta\,,
\eeq
where $\epsilon_{\mu\nu\alpha\beta}$ is the Levi-Civita tensor, and $k_{1,2}$ are the photon four-momenta, see \cref{Tutorial:ALPphotonATeecolliders:FeynmanRule}.

Before jumping into calculations, let us try to build a simplified and approximate picture of what we expect to obtain. The two relevant parameters in the Lagrangian in \cref{eq:Tutorial:ALP-photon:Lagrangian} are the ALP mass, $m_a$, and its coupling to photons, $g_{a\g\g}$. On the experimental side, the two quantities driving the phenomenology of ALP production in $e^+e^-$ collisions and the subsequent ALP detection, are the center of mass energy, $\sqrt{s}$, and the typical detector size, $L_\text{det}$ (we focus on the case where the detector, as in Belle II, surrounds the collision point).

The ALP is produced in the processes of the form $e^+e^-\to X + a$, where $X$ represents any SM particles in the final state. The maximal energy in principle available for ALP production is $\tilde E = \sqrt{s} - m_X$, and is smaller, the larger  the total mass $m_X$ of the SM particles that are also being produced in association with the ALP. For $m_a<\tilde E$, the ALP can be produced on-shell, that is, as a real particle. 
Conversely, for $m_a>\tilde E$ the ALP can only participate in the processes as a virtual particle, contributing to  $e^+e^-\to X + \g + \g$. 
Compared to $e^+e^-\to X + a$ this process is phase space suppressed, and suppressed by additional factor of $g_{a\gamma\gamma}^2$.
We therefore expect the ALP search to be most effective in the regime $m_a<\sqrt{s}-m_X$, while for heavier ALP masses the sensitivity of the experiment gets reduced very quickly. 
We discuss this in greater detail in \cref{subsec:Tutorial:ALPphotonATeecolliders:Production}

If produced on-shell, ALP propagates a distance $\ell_a$ before decaying; approximately, $\ell_a\sim c\tau_a$, where $c$ is the speed of light and $\tau_a$ the ALP lifetime.
The  ALP lifetime $c\tau_a$ depends on ALP mass and its couplings to the SM, see \cref{eq:ALPlifetime:photon,eq:ctau_a:leptons}.  If couplings to photons dominate, then  
$c\tau_a\sim \big( g_{a\g\g}^2m_a^3\big)^{-1}$, ignoring order one factors and factors of $\pi$.  
Light or weakly coupled ALPs travel for a long distance, while heavy or strongly coupled ones have a shorter lifetime, and can even decay promptly, coincident with collision point within experimental resolution. Comparing $c\tau_a$ with detector size $L_{\rm det}$ gives an indication of what signals to expect.
If $c\tau_a \ll L_\text{det}$, the decay $a \to \gamma\gamma$ typically occurs within the detector, leading to a \textit{visible} multiphoton signature.
Conversely, if $c\tau_a \gg L_\text{det}$, the ALPs will typically decay outside the detector volume, resulting in an \textit{invisible} signature due to the missing energy carried away by the undetected ALP. We will explore this behavior in more detail in \cref{subsec:Tutorial:ALPphotonATeecolliders:Decay}.

\subsection{Production}
\label{subsec:Tutorial:ALPphotonATeecolliders:Production}

\begin{figure}[t]
	\centering
    \begin{subfigure}{0.4\linewidth}
    \includegraphics[width=0.9\linewidth]{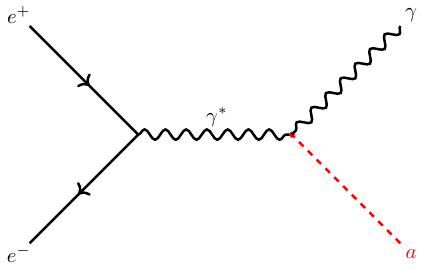}\end{subfigure}
    \hspace{0.4cm}
    \begin{subfigure}{0.4\linewidth}
    \includegraphics[width=0.9\linewidth]{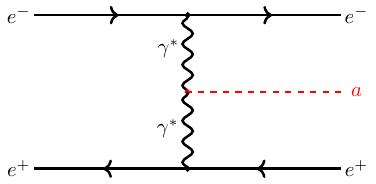}
    \end{subfigure}

    \caption{Feynman diagrams for the ALP production at $e^+e^-$ colliders, assuming only coupling to two photons. }
    \label{Tutorial:ALPphotonATeecolliders:ProductionDiagrams}
\end{figure}

For an ALP that only couples to photons the two main production mechanisms in $e^+e^-$ collisions are shown in \cref{Tutorial:ALPphotonATeecolliders:ProductionDiagrams}: the $2 \to 2$ process (left diagram) is known as "associated production" (or ALP-strahlung), while the $2 \to 3$ process (right diagram) is referred to as "photon fusion".

Let us first focus on the associated ALP production, $e^+e^-\to \gamma a$. The corresponding differential cross section is given by
\beq\label{eq:Tutorial:ALP-photon:TotDiffCrossSec:AssocProd}
\begin{split}
{\rm d}\sigma &= \frac{{\rm d}^3p_\g}{2E_\g(2\pi)^3}\frac{{\rm d}^3p_a}{2E_a(2\pi)^3}\frac{|\bar{\cal M}|^2}{4\sqrt{(p_{e^+}\cdot p_{e^-})^2 - m_e^4}} \\
&\times(2\pi)^4\delta^{(4)}\lp p_{e^+} + p_{e^-} - p_\g - p_a \rp\,,
\end{split}
\eeq
where $|\bar{\cal M}|^2$ is the spin averaged amplitude squared of the process. 
To figure out the kinematics it is useful to work in the center of mass frame, so that ${\vec p}_{e^+} = -{\vec p}_{e^-}$. 
Imposing the energy and momentum conservation, we obtain
\beq
p_\g^2 = 0 = \lp p_{e^+} + p_{e^-} - p_a \rp^2 = s + m_a^2 - 2\sqrt{s}E_a\,,
\eeq
where we used the center of mass frame relation $p_{e^+} + p_{e^-} = (\sqrt{s},0,0,0)$, and the on-shellness condition for the ALP, $p_a^2 = m_a^2$. 
The ALP energy, $E_a$, and the energy of the final state photon, $E_{\g,\text{recoil}}$, are thus
\begin{align}\label{eq:Tutorial:ALP-photon:Ea:AssocProd}
E_a &= \frac{s + m_a^2}{2\sqrt{s}}\,, \\
E_{\g,{\rm recoil}} &= \frac{s - m_a^2}{2\sqrt{s}}\,.\label{eq:Tutorial:ALP-photon:Eg:AssocProd}
\end{align}
Note that the energy of the photon is also equal to the ALP momentum, $|\vec p_a |=E_{\g,\text{recoil}}$.

The total cross section for $e^+e^-\to \gamma a$ is obtained after a straightforward integration, which we leave as an exercise for the reader.\footnote{A useful starting point is to divide the $\delta^{(4)}$ function into spatial and energy components
\beq
\delta^{(4)}\lp p_i - p_f\rp = \delta^{(3)}\lp {\vec p}_i - {\vec p}_f\rp \delta\lp E_i - E_f\rp\,,
\eeq
where $p_{i,f}$ are the total initial and final four-momenta, respectively, and similarly for three-momenta and energies. Alternatively, one of the integrals can be rewritten in the invariant form; for an on-shell particle with four-momentum $p$, energy $E$ and mass $m$, we have
\beq
\int \frac{{\rm d}^3p}{2E(2\pi)^3} = \int \frac{{\rm d}^4p}{(2\pi)^3}\delta\lp p^2 - m^2 \rp \theta\lp E \rp\,,
\eeq
where $\theta$ is the Heaviside function.
} The result is
\beq\label{eq:Tutorial:ALP-photon:ProdCrossSec:AssocProd}
\frac{{\rm d}\sigma}{{\rm d}\cos\theta_{\rm CM}} = \frac{\alpha g_{a\g\g}^2}{32} \lp 1 + \cos^2\theta_{\rm CM} \rp \lp 1 - \frac{m_a^2}{s} \rp^3\,,
\eeq
where $\theta_{\rm CM}$ is the angle between the outgoing ALP and the electron-positron axis in the CM frame. We kept the cross section in differential form in terms of $\theta_\text{CM}$, which will be useful later on, when we will include the effect of limited angular acceptance of the experimental setup. Note the dependence on the ALP mass: for $m_a^2$ well below $s$, the term proportional to the ALP mass quickly disappears, with the cross section practically independent of $m_a$. For $m_a^2\to s$, the cross section quickly drops to zero as $(s-m_a^2)^3\propto ( |\vec p_a|/\sqrt{s})^3$. Here, one power of $|\vec p_a|$ comes from the phase space suppression, and two powers from the fact that this is a $p-$wave process, with $\big|\bar {\cal M}|\propto |\vec p_a|$.

Next, let us consider the photon-fusion ALP production, \cref{Tutorial:ALPphotonATeecolliders:ProductionDiagrams} (right). Since this is a $2\to3$ scattering, the calculation of the cross section becomes more complicated. 
We can appreciate the increased complexity by counting the number of independent momenta and angles parametrizing the final phase space in a fully differential cross section; in other words, by counting how many integrals are required to obtain the total rate.
As a warm up, let us count the number of free parameters describing the $2\to 2$ differential scattering cross section for $e^+e^- \to \gamma a$, \cref{eq:Tutorial:ALP-photon:TotDiffCrossSec:AssocProd}. The kinematics of the $\gamma +a$ final state is desribed by 8 parameters, the 4 components of $p_a$ and the 4 components of $p_\gamma$ four-momenta. These satisfy two on-shell conditions, $p_a^2=m_a^2$ and $p_\gamma^2=0$, reducing the number of free parameters to 6 --- the  $\vec p_a$ and $\vec p_\gamma$ components in the differentials in \cref{eq:Tutorial:ALP-photon:TotDiffCrossSec:AssocProd}. Finally, the energy and momentum conservation (imposed by the $\delta^{(4)}$ function in \cref{eq:Tutorial:ALP-photon:TotDiffCrossSec:AssocProd}) reduces the number of free parameters to a total of 2, 
which in deriving \cref{eq:Tutorial:ALP-photon:ProdCrossSec:AssocProd} we chose to be $E_a$ and $\cos\theta_{\rm CM}$.
Next, let us repeat the counting for a $2\to 3$ process, $e^+e^-\to e^+e^-\gamma$. Adding a third particle in the final state increases the number of free parameters by 4, while there is only one additional constraint, the on-shellness condition for the third particle. We are thus left with 5 free parameters -- or integrals -- to deal with.
While not exceedingly difficult, an analytic solution is still quite cumbersome to obtain. Luckily, we can resort to a numerical simulation/integration, with tools that can compute both the total and differential cross sections readily available. During the tutorial we used {\tt MadGraph v3.5.1}~\cite{Alwall:2014hca}, 
which is quite flexible and as such is very commonly used in particle physics phenomenology community. 

\begin{figure}[t]
	\centering
    \includegraphics[width=.45\linewidth]{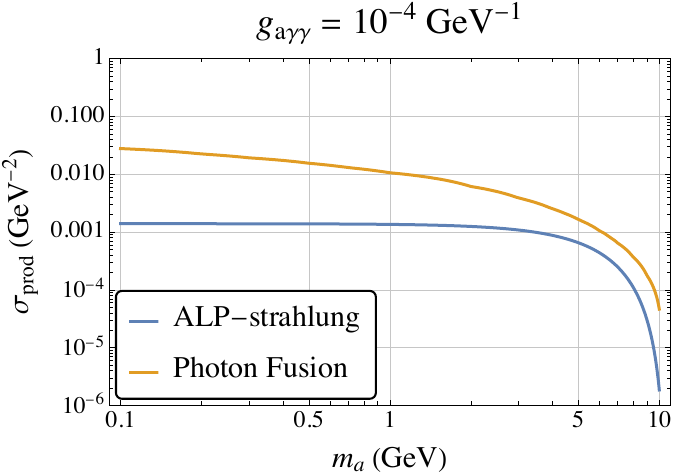}
    \hspace{1cm}
    \includegraphics[width=.44\linewidth]{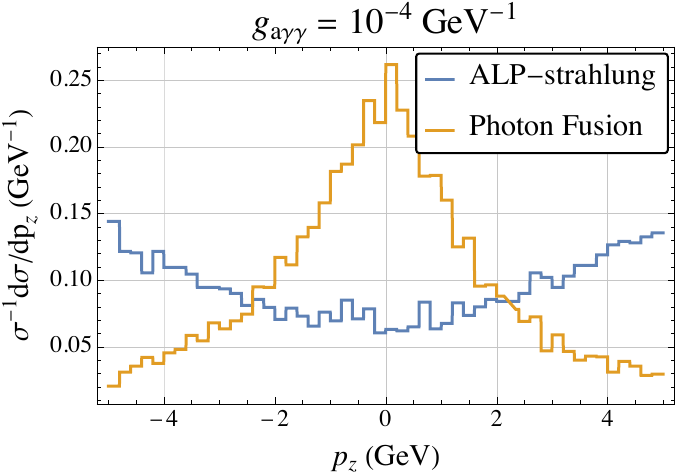}
    \caption{Left: The total cross section for ALP production in $e^+e^-$ collisions at Belle II, via  ALP-strahlung (blue) and photon fusion (orange).  Right: the distribution of ALP momentum component along the beam axis, in the center of mass.  
    }
    \label{Tutorial:ALPphotonATeecolliders:SigmaPlots}
\end{figure}

\Cref{Tutorial:ALPphotonATeecolliders:SigmaPlots} shows numerical results for the case of $g_{a\g\g}=10^{-4}\,\text{GeV}^{-1}$. Left panel shows the results for the total cross section, at the Belle II center of mass energy $\sqrt{s}=10.58\,$GeV, where for all  ALP masses the photon-fusion cross section is found to be larger than the ALP-strahlung. In photon fusion 
the ALP is preferentially  produced at rest, see right panel in \cref{Tutorial:ALPphotonATeecolliders:SigmaPlots}, with the final $e^+$ and $e^-$ mostly continuing in the respective forward directions. This means that the signal of the ALP production will consist of two back-to-back photons from the ALP decay, and the missing energy due to the very forward $e^+e^-$ that are lost along the beam pipe, which is not instrumented.  
Unfortunately, this signature is plagued by large beam-induced backgrounds, especially for low-mass ALPs. Despite its larger cross section we will thus ignore the photon fusion production, and focus in the rest of the tutorial on the associated ALP production and the resulting signatures.

\subsection{ALP decays}
\label{subsec:Tutorial:ALPphotonATeecolliders:Decay}
For an ALP coupling predominantly to photons the main decay channel is $a\to\g\g$. The corresponding decay width, $\Gamma(a\to\g\g)$,  in the notation of \cref{sec:ALPdecays}, was given in \cref{eq:ALP-widths}. Let us now re-calculate it,  using our alternative notation for ALP--photon couplings, \cref{eq:Tutorial:ALP-photon:Lagrangian}, and show the computational steps in detail.
First of all, it is convenient to work in the ALP rest frame, in which $p_a = (m_a,0,0,0)$. The $a\to 2\gamma$ is a 2-body decay, and the two daughter particles are massless; the only relevant energy scale is thus the ALP mass $m_a$. Conservation of energy and the symmetry of the problem imply that the available energy, $m_a$, is shared equally among the two photons, giving $E_{\g_1} = E_{\g_2} = m_a/2$. The invariant mass of the two photon system is thus
\beq\label{eq:Tutorial:ALP-photon:Mgg}
M_{\g\g}^2 = (k_1 + k_2)^2 = m_a^2\,.
\eeq
In the lab frame, the conservation of linear momentum implies that the three particles, $a$ and $2\gamma$, lie in a plane. In the ALP rest frame, the conservation of linear momentum thus implies that the two photons are back-to-back, ${\vec k}_1 = -{\vec k}_2$; in the ALP rest frame the angle between the two photons is $\pi$. Dimensional analysis gives us the dependence of $\Gamma(a\to\g\g)$ on $m_a$. The decay width $\Gamma(a\to\g\g)$ has mass dimension $1$, and is proportional to the coupling constant squared, $g_{a\g\g}^2$, since the decay width is proportional to the amplitude squared. The mass dimension $-2$ of $g_{a\g\g}^2$ needs to be compensated by $m_a^3$ to obtain the right dimensionality of $\Gamma(a\to\g\g)$. Including all the remaining factors gives
\beq\label{eq:Tutorial:ALP-photon:DecayWidth}
\Gamma(a\to\g\g) = \lp \frac{4\pi}{32\pi^2} \rp \times \lp \frac{g_{a\g\g}}{4} \rp^2 \times m_a^3 = \frac{g_{a\g\g}^2m_a^3}{128\pi}\,.
\eeq
The $1/(32\pi^2)$ prefactor comes from the combination of the density of states in the two-body phase space and the definition of the decay width as a transition rate probability, while the $4\pi$ factor in the numerator is due to the integration over the solid angle, describing the orientation of the two photons in the final state (this integration is trivial -- there is no angular dependence). We leave it as an exercise for the reader to compute the above decay width rigorously, making use of \cref{eq:Tutorial:ALP-photon:FeynRule} (also the PDG kinematics review, Ref.~\cite{ParticleDataGroup:2022pth}, is a very helpful resource).

\begin{figure}[t]
	\centering
    \includegraphics[width=0.5\linewidth]{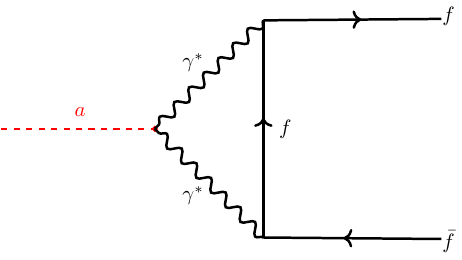}
    \caption{Feynman diagrams for the ALP decay at  one-loop.}
    \label{Tutorial:ALPphotonATeecolliders:DecayDiagrams}
\end{figure}

Even if coupling to photons is assumed to be the only one at tree level, others will be generated at loop level. 
By closing the loop with a fermion line, as shown in \cref{Tutorial:ALPphotonATeecolliders:DecayDiagrams}, ALP can decay to two charged leptons or to quarks (hadrons), if kinematically allowed. We can estimate that these decay widths will be suppressed by $\lp \alpha/4\pi \rp^2\sim10^{-7}$ with respect to \cref{eq:Tutorial:ALP-photon:DecayWidth}: the two additional QED vertices introduce $e^2$ in the diagram, and the loop comes with a $1/(4\pi)^2$ suppression factor. The opposite is also true, if ALP couples only to leptons or quarks at tree level, this will generate ALP-photon couplings at one loop, with the result given in \cref{eq:cgamgameff}. 

In the rest of the tutorial we will assume that the ALP-photon coupling dominates, and ignore any loop induced decays to leptons and hadrons. 
In this limit, the total decay width of the ALP is simply given by the two-photon channel,
\beq
\Gamma_a^{\rm tot}\equiv\Gamma(a\to\g\g)\equiv\Gamma_a.
\eeq
The \textit{proper lifetime} of the ALP, i.e., the lifetime as measured in the ALP rest frame, is then (setting $c=\hbar=1$ as in most of these lecture notes)
\beq
\tau_a \equiv \Gamma_a^{-1} = \frac{128\pi}{g_{a\g\g}^2m_a^3}\,.
\eeq
The ALP, however, can be produced with a possibly relativistic energy $E_a = (s + m_a^2)/(2\sqrt{s})$, see \cref{subsec:Tutorial:ALPphotonATeecolliders:Production}. We therefore need to take into account relativistic time dilatation. The lifetime of a fast moving particle can be significantly longer than its proper lifetime, and is equal to $\tau = \g_a\tau_a$, where $\g_a = E_a/m_a$ is the ALP Lorentz factor and $\beta = |{\vec p}_a|/E_a$ its velocity in units of $c$.
The \textit{boosted decay length} of an ALP is thus
\beq\label{eq:Tutorial:ALP-photon:DecayLength}
\ell_a = \g\beta c\tau_a = \frac{|{\vec p}_a|}{m_a}\times c \times \frac{128\pi}{g_{a\g\g}^2m_a^3}\,.
\eeq

It is instructive to consider a realistic numerical example.  
Assume a collider with $\sqrt{s}\sim10$ GeV, and an ALP with mass light enough that we can approximate $|{\vec p}_a|\sim E_a$ and $E_a \sim \sqrt{s}/2$. \Cref{eq:Tutorial:ALP-photon:DecayLength} then gives
\beq
\ell_a\simeq 0.4~{\rm cm}\times\lp \frac{g_{a\g\g}}{10^{-5}~\GeV^{-1}} \rp^{-2} \times\, \biggr( \frac{m_a}{\GeV} \biggr)^{-4}\times \biggr(\frac{\sqrt{s}}{10\,\text{GeV}}\biggr)\,.
\eeq
Note the quadratic scaling with the coupling: the average distance traveled by an ALP with $m_a = 1$ GeV will go from tens of microns to half a meter, if we change $g_{a\g\g}$ from $10^{-4}\,\text{GeV}^{-1}$ to $10^{-6}\,\text{GeV}^{-1}$. The change with $m_a$ is even faster, since $\ell_a\propto m_a^{-4}$. Reducing the mass of $m_a$ well below a GeV, one quickly runs out of having any ALPs decaying within the detector volume. 

For highly boosted ALPs there is another experimentally important effect: 
for small ALP masses (large boosts), the final two photons may only have a small angular separation, $\Delta\theta_{\g\g}$, in the lab frame.
In order to experimentally observe two separate photons, $\Delta\theta_{\g\g}$ needs to be larger than the angular resolution $\Delta\theta_{\rm res}$ of the detector. For smaller $\Delta\theta_{\g\g}$ the two electromagnetic showers initiated by the two $a\to \gamma\gamma$ photons merge into a single one.  

To calculate $\Delta\theta_{\g\g}$, we use the Lorentz transformation that takes us from the ALP rest frame to the lab frame. Let us choose the $z$-axis to be in the direction of the ALP momentum in the lab frame, $\vec p_a$, while the two photons are taken to lie in the plane formed by the $z-$ and $y$-axes. Indicating with primes the quantities in the rest frame, we have
\beq
\begin{split}
k_y &= k\sin\theta \equiv k_y^\prime = k'\sin\theta^\prime\,, \\
k_z &= k\cos\theta \equiv \g_a \lp k'\cos\theta^\prime + \beta_a k' \rp\,,
\end{split}
\eeq
where the photon momentum in the rest (lab) frame,  $k\equiv|\vec k|$ ($k'\equiv|\vec k'|)$, is at an angle $\theta$ ($\theta'$) with respect to the $z$-axis. 
For the lab frame angle $\theta$ we thus have
\beq\label{eq:Tutorial:ALP-photon:Boosting}
\tan\theta = \frac{1}{\g_a}\frac{k'\sin\theta^\prime}{k'\cos\theta^\prime + \beta_a k^\prime} \xrightarrow{\theta'=\pi/2, \beta_a=1} \frac{1}{\gamma_a}\,,
\eeq
where on the r.h.s. we took the ultra-relativistic limit $m_a\ll\sqrt{s}$ and chose a photon moving in the $y$-direction in the ALP rest frame, $\theta'=\pi/2$, as a representative example. Taking into account that the direction of the other photon in the rest frame is $-(\pi-\theta')$, this representative kinematics in fact gives the smallest angular separation $\Delta \theta_{\gamma\gamma}$ in the lab frame, so that 
\beq
\label{eq:Deltatheta:gg}
\Delta\theta_{\g\g}\gtrsim \frac{2}{\gamma_a},
\eeq
where the typical value is of the order of this lower bound. Here, $\gamma_a\simeq \sqrt{s}/2m_a$ for symmetric $e^+e^-$ collisions (Belle II has somewhat asymmetric $e^+$ and $e^-$ beams of 4\,GeV and 7\,GeV so that this estimate still applies at an ${\mathcal O}(1)$ level).

\subsection{Types of signals}
\label{subsec:Tutorial:ALPphotonATeecolliders:SignalTypes}
We are now ready to explore the types of signals we can expect for photo-philic ALPs produced in $e^+e^-$ collisions.
To summarize; we have an ALP produced on-shell, thus with mass $m_a<\sqrt{s}$, via the associated production process, $e^+e^-\to\g+a$; the energy of the ALP is fixed from the kinematics to be $E_a = (s + m_a^2)/{2\sqrt{s}}$, while the recoiling photon's energy is $E_{\g,{\rm recoil}} = (s - m_a^2)/{2\sqrt{s}}$, see \cref{eq:Tutorial:ALP-photon:Ea:AssocProd,eq:Tutorial:ALP-photon:Eg:AssocProd}. The ALP travels a distance $\ell_a$ before decaying into two photons, $a\to \gamma\gamma$, see \cref{eq:Tutorial:ALP-photon:DecayLength}. The invariant mass of the two photon system is $M_{\g\g} = m_a^2$, see \cref{eq:Tutorial:ALP-photon:Mgg}, where in the lab frame the two daugther photons have an angular separation $\Delta\theta_{\g\g}\sim{\mathcal O}(4m_a/\sqrt{s})$, see \cref{eq:Deltatheta:gg}.

\begin{figure}[t]
	\centering
	\includegraphics[width=0.8\linewidth]{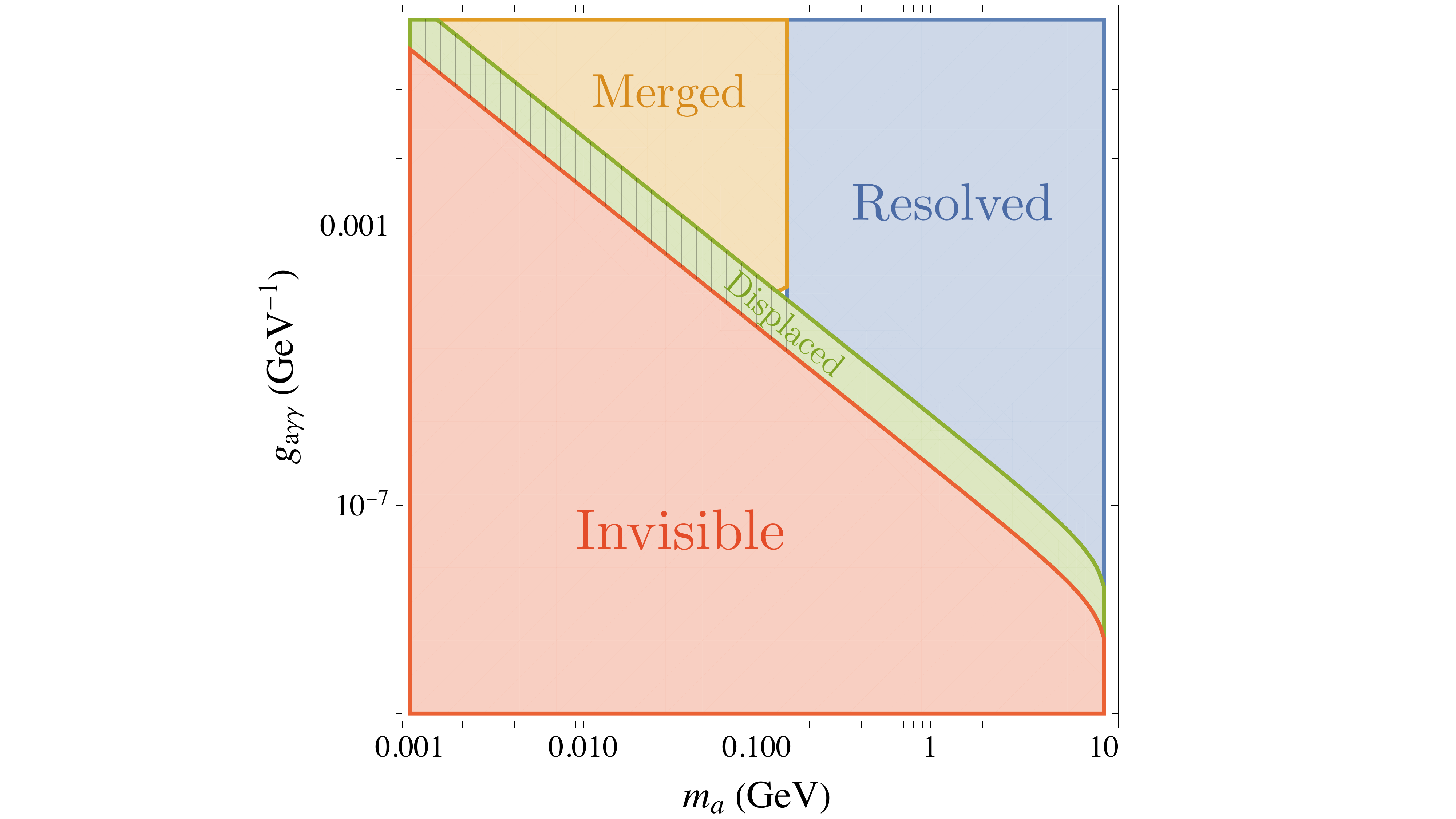}
    \caption{Different types of signals expected at Belle II in the parameter space spanned by $m_a$ and $g_{a\g\g}$. The hatched region indicates where the two photons in the displaced event cannot be resolved and become merged, see main text for details. Adapted from Ref.~\cite{Dolan:2017osp}. }
    \label{Tutorial:ALPphotonATeecolliders:SignalTypes:BelleII}
\end{figure}

Out of these ingredients four distinct types of signal can emerge, organized according to where the ALP decays and how boosted the ALP is (see also \cref{Tutorial:ALPphotonATeecolliders:SignalTypes:BelleII}):
\begin{itemize}
    \item \textbf{Visible: prompt and resolved.} This requires ALP to decay within the inner part of the detector, within the detector resolution for the collision vertex,  $\ell_a\lesssim L_{\rm min}$, and to have the angle between the two $a\to \gamma\gamma$ photons to be large enough $\Delta\theta_{\g\g}\gtrsim\Delta\theta_{\rm res}$. The signal consists of three photon induced electromagnetic showers in the detector, pointing to the $e^+e^-$ interaction point (IP).
    \item \textbf{Visible: displaced and resolved.} This occurs for $L_{\rm min}\lesssim \ell_a\lesssim L_{\rm max}$ and $\Delta\theta_{\g\g}\gtrsim\Delta\theta_{\rm res}$; the signal consists of three photon-induced electromagnetic showers in the detector, one of which is pointing to the $e^+e^-$ IP, while the other two point to a vertex that is displaced from the IP.
    \item \textbf{Visible: merged.} This occurs for $\ell_a\lesssim L_{\rm max}$ and $\Delta\theta_{\g\g}\lesssim\Delta\theta_{\rm res}$, so that the two photons from $a\to \gamma\gamma$ are merged. The signal consists of two electromagnetic showers in the detector, pointing to the $e^+e^-$ IP. Note that even if $\ell_a\gtrsim  L_{\rm min}$ and so the ALP decay is displaced, this is not observable since the photons from $a\to \gamma\gamma$ decay are merged (one can still potentially distinguish this signature from a true prompt decay, if the decay $a\to \gamma\gamma$ occurs behind the electromagnetic calorimeter, e.g., in muon chambers --- this now gets us into the details of the particular particle physics detector design.)
    \item \textbf{Invisible.} This requires $\ell_a\gtrsim L_{\rm max}$; the signal consists of a single photon-induced electromagnetic shower in the detector, pointing to the $e^+e^-$ IP, and the missing energy.
\end{itemize}

Above, we indicated with $L_{\rm min}$ the detector vertex resolution (in this case, from its electromagnetic calorimeter) and $L_{\rm max}$ the typical size of the detector. Note that this is a rough approximation, first, because detectors are not spherical, and secondly, because the probability distribution of a decaying particle follows an exponential law. Each signature is relevant for different parts of the ALP parameter space, depending on the experimental setup, and requires a different search strategy.

\subsection{Short detour - the $\chi^2$ statistics}
\label{subsec:Tutorial:ALPphotonATeecolliders:chi2}
Before proceeding, let us do a short detour into statistics. This will then prove useful when computing bounds on ALP parameters. 
We will focus on the Maximum Likelihood Estimate (MLE), however, a more complete discussion can be found in the "Statistics" section of the PDG review~\cite{ParticleDataGroup:2022pth}.

Let us consider $N$ independent measurements $y_i$ performed at kinematical points $x_i$ (in our example the $y_i$ could be the number of ALP decays in different  bins of kinematics variables $x_i$, such as scattering angles, or di-photon invariant mass). The likelihood function for a set of model parameters $\theta$, $L(\theta) = p(y(x)|\theta)$, can be written as
\beq
L(\theta) = p(y(x)|\theta) = \prod_i^N f(y_i(x_i)|\theta)\,,
\eeq
where $f$ are probability distributions for $y_i$. 
In our case, $\theta$ can be the ALP mass and the coupling to photons.
If the $y_i$ are Gaussian distributed independent random variables, then\footnote{We do not display the normalization factor of the Gaussian distribution, since it is not relevant for the discussion below.}  
\beq
\label{eq:fyixi}
f(y_i(x_i)|\theta) \propto \exp\left[-\frac{(y_i -\mu(x_i,\theta))^2}{2\sigma_i^2} \right]\,,
\eeq
where $\mu$ and $\sigma$ are the mean and the standard deviation, respectively.

Computationally it is more efficient to consider the log of the likelihood function, as it turns products into sums and removes the exponentials. The log-likelihood is thus given by
\beq\label{eq:Tutorial:ALP-photon:Chi2Function}
-2\ln L(\theta) =  \chi^2(\theta) + {\rm const.}\,,
\eeq
where the function
\beq\label{eq:Tutorial:ALP-photon:Chi2Function:2}
\chi^2(\theta) = \sum_i^N \frac{(y_i -\mu(x_i,\theta))^2}{\sigma_i^2}\,,
\eeq
is a random variable whose values follow the $\chi^2$ probability distribution. 
As the estimator $\hat\theta$ for the true $\theta$ parameter values we use $\theta$ that maximizes the log-likelihood function, or equivalently, that minimizes the $\chi^2(\theta)$ function. Note that this procedure is, in the above example, equivalent to the least squares method.

The MLE can also be used as a means of a hypothesis test. The Neyman-Pearson lemma states that the most powerful hypothesis test follows the likelihood ratio 
\beq
\lambda(y) = \frac{p(y(x)|\hat\theta_1)}{p(y(x)|\hat\theta_0)} = \frac{L(\hat\theta_1)}{L(\hat\theta_0)}\,.
\eeq
Here, $\hat \theta_0$ and $\hat \theta_1$ define the two different hypotheses that are being tested. For Gaussian distributed events then
\beq
-2\ln\lambda = -2 \ln L(\hat\theta_1) + 2 \ln L(\hat\theta_0) = \chi^2(\hat\theta_1) - \chi^2(\hat\theta_0) \equiv \Delta\chi^2\,.
\eeq
It can then be shown that 
\beq\label{eq:Tutorial:ALP-photon:DeltaChi2}
\Delta\chi^2 = F_{\chi_m^2}^{-1}(1-p)\,, 
\eeq
where $F_{\chi_m^2}^{-1}$ is the chi-square quantile of the $\chi^2$ distribution with $m$ degrees of freedom,\footnote{Not to be confused with the $\chi^2$ function defined in \cref{eq:Tutorial:ALP-photon:Chi2Function}.} and $1-p$ the coverage percentage of the data sample. That is, each value of $\Delta\chi^2$ gives the probability $p$ of falsely rejecting the null hypothesis, that is, whether $\hat \theta_1$ is sampled from the same probability distribution as $\hat \theta_0$. 
We can use the $\Delta \chi^2$ also to quantify the \textit{goodness of fit}, that is how well a hypothetical value of $\theta$ describes experimental data. First one finds the value $\hat \theta_0$ that minimizes the $\chi^2(\theta)$. The difference $\Delta \chi^2(\theta)=\chi^2(\theta)-\chi^2(\hat \theta_0)$ is then $\chi^2$ distributed for Gaussian errors, and thus can be used to set allowed $p-$value confidence level intervals on $\theta$, see \cref{eq:Tutorial:ALP-photon:DeltaChi2}.

\paragraph{How do we use the above to set bounds on ALP couplings?} For the case of ALP production at Belle-II the $y_i$ in \cref{eq:fyixi} are the number of of events $N_i$ in bins of two-photon invariant mass, $M_{\g\g}$ (these are the $x_i$ in \cref{eq:fyixi}, one could have equally used bins of the recoiling photon's energy, $E_{\g,{\rm recoil}}$, see \cref{eq:Tutorial:ALP-photon:Eg:AssocProd}). For each bin let us assume that the $N_i$ are large enough such that the central limit theorem applies, and thus $N_i$ are Gaussian distributed, with statistical uncertainty given by $\sigma_i=\sqrt{N_i}$ (for simplicity we neglect systematic uncertainties). 

For our example, the parameters $\theta$ consist of all the parameters of the SM, $\theta_{\rm SM}$, and in addition, the ALP mass and its coupling to photons, $\theta = \{ \theta_{\rm SM}, m_a, g_{a\g\g} \}$. The null-hypothesis is $\theta_0 = \{ \theta_{\rm SM}, 0,0 \}$, i.e., that the observed events are entirely due to the SM processes; the alternative hypothesis is $\theta_1 = \{ \theta_{\rm SM}, m_a^*, g_{a\g\g}^* \}$, i.e., that there is in addition a signal from ALP production and decay, characterized by $\{m_a^*, g_{a\g\g}^* \}$. Let us next assume that data agree well with the SM predictions. We can than proceed to set limits on $g_{a\g\g}$ as a function of ALP mass using 
\beq\label{eq:Tutorial:ALP-photon:DeltaChi2:Events}
\Delta\chi^2 = \sum_i \left[ \lp \frac{N_{{\rm meas},i} - (N_{B,i} + N_{S,i}(m_a,g_{a\g\g}))}{\sigma_i} \rp^2 - \lp \frac{N_{{\rm meas},i} - N_{B,i} }{\sigma_i} \rp^2 \right]\,,
\eeq
where $N_{{\rm meas},i}$ ($N_{B,i}$,$N_{S,i}$) are the number of measured (expected background, signal) events in $i$-th bin, where for simplicity we assumed that $\chi^2$ is minized for $g_{a\gamma\gamma}=0$ (the so called Asimov set). The errors in each bin are taken to be statistical only, $\sigma_i = \sqrt{N_{{\rm meas},i}}$.
For a fixed value of ALP mass, $m_a^*$, the $1-p=90\%$ confidence level (CL) bound on  $g_{a\gamma\gamma}$ is thus given by the value $g_{a\gamma\gamma}|_{0.9}$ that satisfies
\beq
\Delta\chi^2(m_a^*,g_{a\g\g}|_{0.9}) = F_{\chi_1^2}^{-1}(0.9) = 2.71\,.
\eeq

As a final working case, let's consider a type of signal which is predicted to be zero in the SM; we expect no background events, $N_{B,i} = 0$. In the limit of no systematic errors, we therefore expect $N_{{\rm meas},i} = 0$. 
As a consequence, any measured event would be a signal of new physics. In this case, where we deal with only a small number of events, a proper treatment would require switching from Gaussian to Poisson distributed data, which we will not do in this tutorial. A simple rule of thumb commonly used to gauge the sensitivity of experiments without detailed estimates of possible backgrounds is to require the observation of at least 3 events to claim a discovery, namely
\beq\label{eq:Tutorial:ALP-photon:DeltaChi2:ZeroBkg90}
N_{\rm S}^{\rm tot}(m_a,g_{a\g\g}) = 3\,.
\eeq

\subsection{Experimental setup - Belle II at Super-KEKB}
\label{subsec:Tutorial:ALPphotonATeecolliders:BelleII}
The Super-KEKB is an electron-positron collider located in Tsukuba, Japan, with a nominal center of mass energy of $\sqrt{s} = 10.58$ GeV.\footnote{Note that the two beams are asymmetrical: the electron have $E_{e^-}\sim7$ GeV, while the positrons $E_{e^+}\sim4$ GeV. This is needed in order to obtain a small boost of the final state particles along the beam direction, $\g\beta=0.28$. We do not include the effects due to asymmetry in our simplified treatment.} The main reason for this choice of energy is to produce the resonance $\Upsilon(4S)$, which  predominantly decays into pairs of $B$ mesons; 
this type of a collider is thus commonly referred to as a $B$-factory. 
After the completion of Belle-II  program the expected integrated luminosity is ${\cal L}_{\rm full} = 50~{\rm ab}^{-1}$, which would correspond to $N_{BB}\sim5\times10^{10}$ created $B^+B^-$ pairs (and same for $B^0{\bar B}^0$ pairs). So far, the ALP searches at Belle-II have been carried out with a significantly smaller collected luminosity of ${\cal L} = 445~{\rm pb}^{-1}$~\cite{Belle-II:2020jti}.

The IP is located centrally, on the symmetry axis of the Belle II detector~\cite{Belle-II:2010dht}. Out of several detector components, we are most interested in the Electromagnetic Calorimeter (ECL), given that our final state contains only photons and electrons, and the Silicon Vertex Detector (SVD). Schematically, the ECL can be thought of as a hollow cylinder, with the IP at its center. The ECL barell is 3m long, with an inner radius of 1.25m, and outer one of 1.55m; the latter is the value of $L_{\rm max}$ considered in our analysis. Similarly, the SVD is positioned close to the IP, with the main purpose to measure the decay vertices of the particles produced at the IP. Its outer radius is 0.14m, which we take to be $L_{\rm min}$ in our rough estimates.

The angular coverage of this setup is close to the full solid angle, but not quite, since particles traveling along the beam line (which is not instrumented) are lost. The Belle II detector has coverage in the range $[12.4\degree,155.1\degree]$ for the polar angle in the CM frame, although the region with the best resolution is found to be $[37.3\degree,123\degree]$\footnote{These values indicate the lab frame acceptance, which need to be boosted in the CM frame as in \cref{eq:Tutorial:ALP-photon:Boosting}, using $\g\beta=0.28$, giving the range $[24.1\degree, 133.4\degree]$.}~\cite{Belle-II:2020jti}.
Finally, the minimum energy required for a track to be detected in the ECL is $E_\g=0.25$ GeV, with an energy resolution of $\Delta E \simeq 2\%$ and a minimum angular resolution to distinguish two tracks of $\Delta\theta_{\rm res} \simeq 0.8\degree$.

A summary of the expected signals at Belle II, in the parameter space of the ALP, is shown in \cref{Tutorial:ALPphotonATeecolliders:SignalTypes:BelleII}. Here we will overlook details related to the vertexing capabilities of the detectors, and as such we will only compute the reach for ``prompt resolved'' and ``invisible'' signatures.

\subsection{Experimental reach - the invisible signature}
\label{subsec:Tutorial:ALPphotonATeecolliders:Invisible}
We expect an invisible signature to be relevant for low mass  ALPs (and thus large boosts) and/or weak couplings, translating into long decay lengths. Note, however, that an unstable particle can still have an exponentially suppressed probability to decay within the detector, which needs to be taken into account.
The survival probability for ALP not to decay within length $L$ from IP is 
\beq
P(L, m_a, g_{a\g\g}) = \exp\lp -\frac{L}{\ell_a(m_a, g_{a\g\g})} \rp\,,
\eeq
with $\ell_a$ given in \cref{eq:Tutorial:ALP-photon:DecayLength}. For $\ell_a\gg L$, the particle is effectively stable on collider length scales, and $P(L)\to1$. For $\ell_a\ll L$ instead, the particle decays rapidly, and $P(L)\to0$.

For ALP production, the total rate of expected invisible events in the Belle II detector is given by
\beq\label{eq:Tutorial:ALP-photon:NumberOfInvEvents}
N_{\rm inv}({\cal L}, m_a, g_{a\g\g}) = {\cal L}\times\sigma_{\rm prod}(m_a, g_{a\g\g})\times P(L_{\rm max}, m_a, g_{a\g\g})\,,
\eeq
where ${\cal L}$ is the collected integrated luminosity, and $\sigma_{\rm prod}$ the total ALP production cross section.
In the center of mass the signal events appear as a single photon of fixed energy, $E_{\g,\rm recoil}$, see \cref{eq:Tutorial:ALP-photon:Eg:AssocProd}. The background events from the SM that can lead to the same signature are highly improbable. For instance, the SM process $e^+e^-\to e^+e^-\g$ has a large cross section, but can fake the mono-photon signal only if both $e^+$ and $e^-$ do not trigger a response in the detector. 

\begin{figure}[t]
	\centering
	\includegraphics[width=0.75\linewidth]{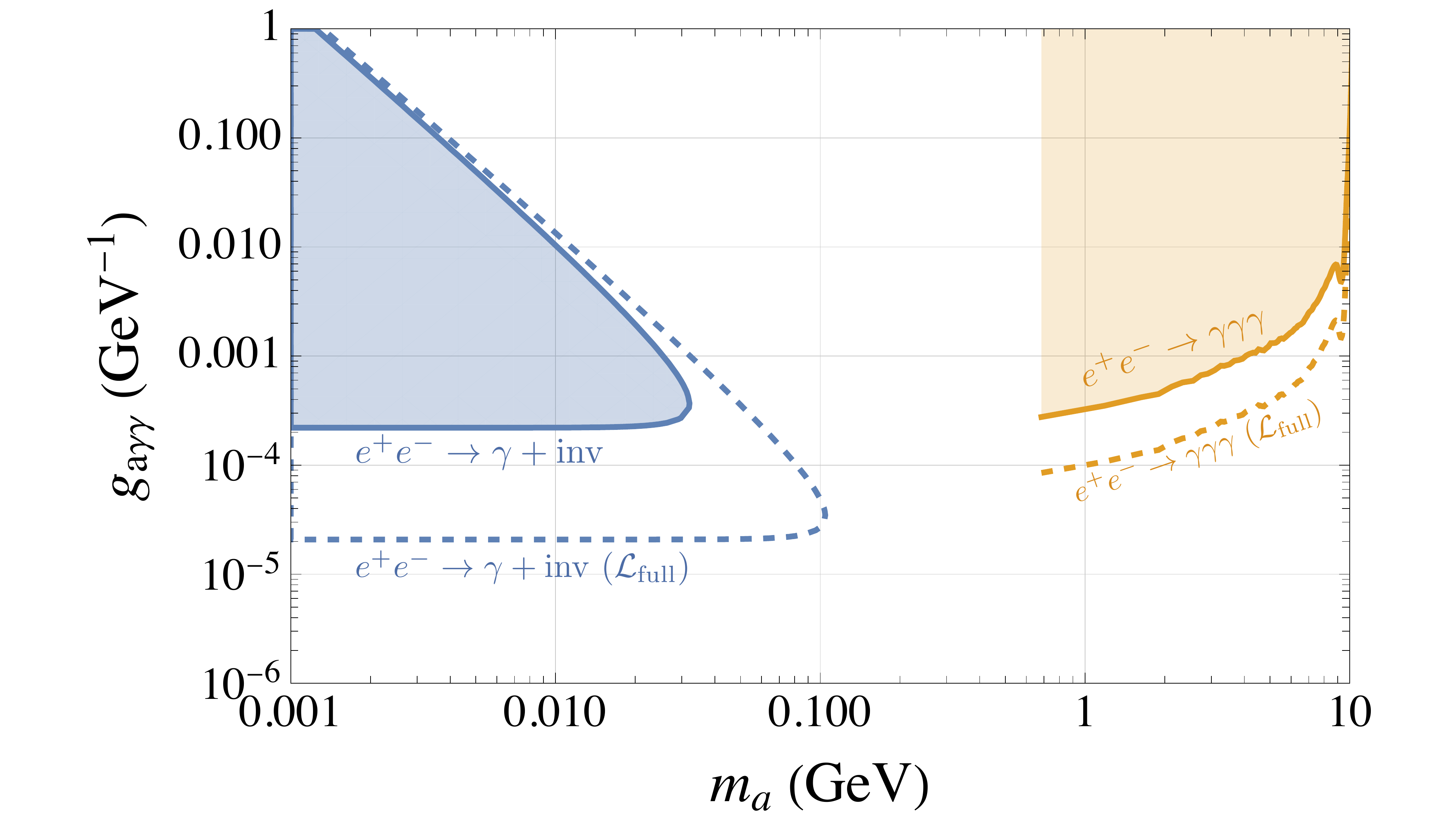}
    \caption{Constraints in the parameters space of $m_a$ and $g_{a\g\g}$, from the invisible (blue) and resolved (orange) signatures, respectively. The dashed lines indicate the results rescaled to the full luminosity expected at Belle II.}
    \label{Tutorial:ALPphotonATeecolliders:BelleII:Bounds}
\end{figure}

Let us thus take the poor man's approach and neglect completely the backgrounds. 
The reach of the experiment, at 90 CL\%, is then set by solving \cref{eq:Tutorial:ALP-photon:DeltaChi2:ZeroBkg90} for $g_{a\g\g}$ at different masses $m_a$.
The results are shown as a blue region in \cref{Tutorial:ALPphotonATeecolliders:BelleII:Bounds} for ${\cal L} = 445~{\rm pb}^{-1}$.
Note the shape of the region probed by Belle II. The upper limit of the excluded region follows roughly the expectation from \cref{Tutorial:ALPphotonATeecolliders:SignalTypes:BelleII}, as ALPs with large enough couplings either decay promptly or displaced, but within the detector. Similarly, heavier ALPs have shorter decay lengths, thus the reach of searches for the invisible signature is also limited in the case of larger $m_a$ values. 

The lower boundary of the excluded region is instead determined by the detector size and the integrated luminosity. For very small couplings, the decay length increases and the survival probability quickly drops to zero. For a given detector with fixed size, the only way to improve the reach is to increase the integrated luminosity, see \cref{eq:Tutorial:ALP-photon:NumberOfInvEvents}. The expected reach for the full projected Belle-II luminosity, ${\cal L} = 50~{\rm ab}^{-1}$, is shown as a dashed blue line in \cref{Tutorial:ALPphotonATeecolliders:BelleII:Bounds}.

\subsection{Experimental reach -- the resolved signature}
\label{subsec:Tutorial:ALPphotonATeecolliders:Resolved}
We expect the resolved signature, $e^+e^-\to\g (a\to\g\g)$, to be relevant for the high-mass part of the available ALP parameter space, where the ALP decays promptly but with small enough boost so that the two photons can still be resolved in the detector.

\begin{figure}[t]
	\centering
	\includegraphics[width=.42\linewidth]{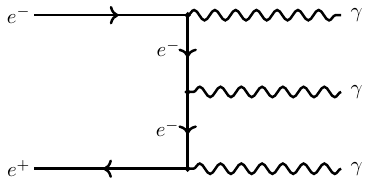}
    \caption{Feynman diagram for the main background process to the resolved photophilic ALP signature.}
    \label{Tutorial:ALPphotonATeecolliders:BelleII:3gBkg}
\end{figure}

Among the photons produced in the resolved signature case, $e^+e^-\to\g a,~a\to\g\g$, the first photon is produced in the center of mass with a fixed energy $E_{\g,{\rm recoil}}$ that depends on the ALP mass, see \cref{eq:Tutorial:ALP-photon:Eg:AssocProd}. The other two photons would form a peak at the invariant mass $M_{\g\g}=m_a$, see \cref{eq:Tutorial:ALP-photon:Mgg}. The signal (when boosted into the CM) will thus consist of a narrow peak in the reconstructed invariant mass of two photons, accompanied by a mono-energetic photon. 
The width of these peaks is determined by the energy resolution of the detector. The expected rate of the events is
\beq
N_{\rm res}(M_{\g\g}) = {\cal L}\times\sigma_{\rm prod}(m_a,g_{a\g\g})\times\epsilon(M_{\g\g})\,,
\eeq
where $\epsilon(m_a)$ indicates the average efficiency of the detector for this particular signal. 

The main difference with respect to the invisible case is that there are large backgrounds due to the SM processes; any channel that results in 3 photons (or electrons that are mis-reconstructed as photons) can fake the ALP signal. The largest background is expected to be due to the QED process $e^+e^-\to\g\g\g$, see \cref{Tutorial:ALPphotonATeecolliders:BelleII:3gBkg}.\footnote{Other, less imporant, sources of backgrounds have also been studied in the actual experimental analysis, see Ref.~\cite{Belle-II:2020jti} for details.} Similarly to the photon-fusion production, it is faster to compute this cross section numerically, e.g., with {\tt MadGraph}, than it is to calculate the analytical expression. Once done, a signal hypothesis can be added to the background to compare with measured data, and extract bounds by means of \cref{eq:Tutorial:ALP-photon:DeltaChi2:Events}. 

\begin{figure}[t]
	\centering
	\includegraphics[width=0.8\linewidth]{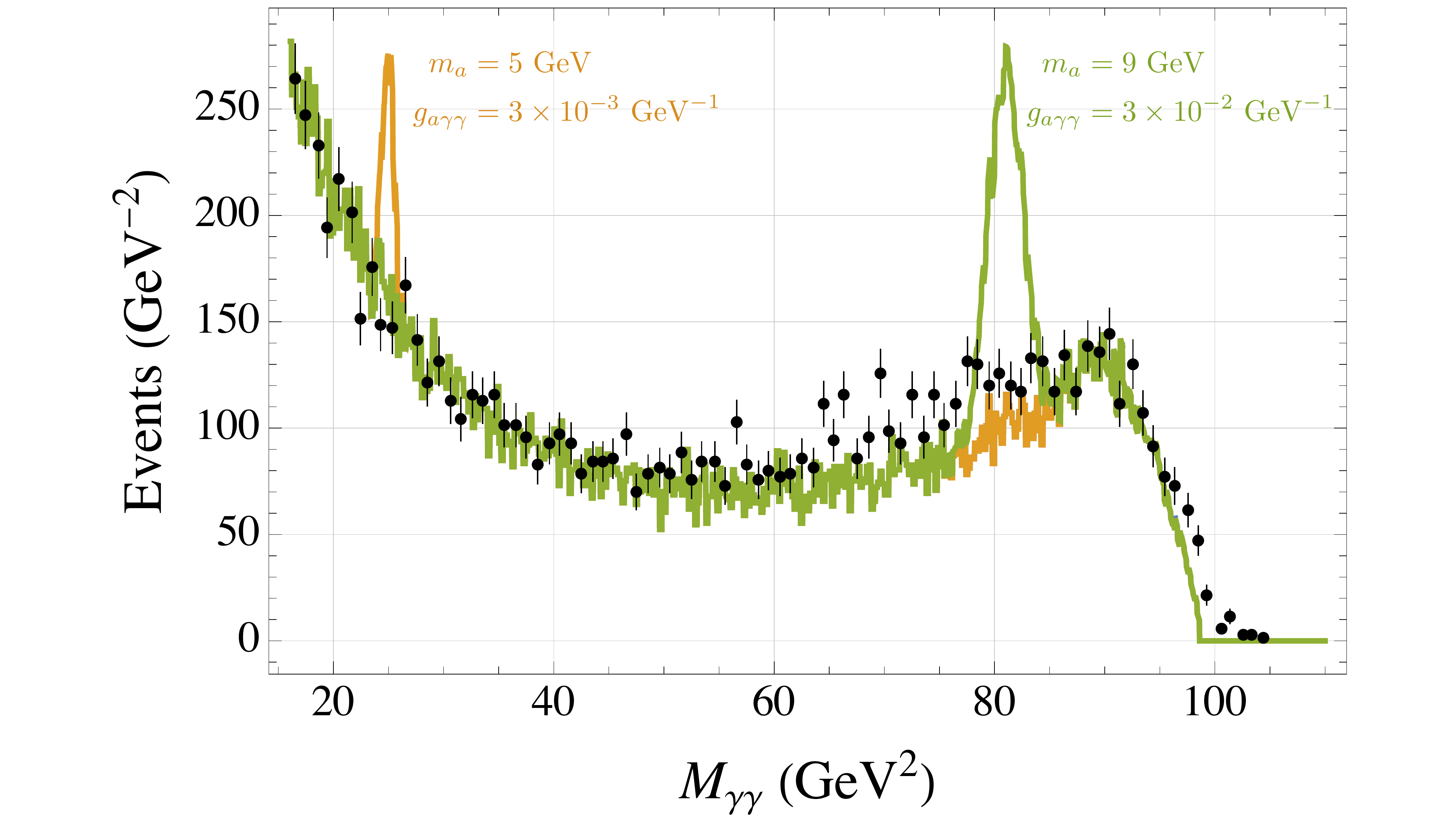}
    \caption{Data points measured at Belle II (black dots), compared with two simulations of the ALP signal. }
    \label{Tutorial:ALPphotonATeecolliders:BelleII:3gDataBkg}
\end{figure}

In \cref{Tutorial:ALPphotonATeecolliders:BelleII:3gDataBkg} we show data from Ref.~\cite{Belle-II:2020jti}, shown as black dots with error bars, as well as  two different signal hypotheses, added on top of the SM background: an ALP of mass $m_a = 5$ GeV and $g_{a\g\g} = 3\times10^{-3}~{\rm GeV}^{-1}$ (green line), and an ALP of mass $m_a = 9$ GeV and $g_{a\g\g} = 3\times10^{-2}~{\rm GeV}^{-1}$ (orange line). In order to properly estimate the statistical significance of the agreement between data and the SM predictions, more refined simulations of the backgrounds are required than our naive treatment, which we will not attempt in this tutorial. 

Instead, we can still arrive at a fair estimate of the bounds on the ALP parameter space by assuming that the measured data-points perfectly describe the expected SM background.  
That is, we take $N_{\rm meas,i} = N_{B,i}$ for all bins. Furthemore we also assume that the measurements are statistically limited, so that $\sigma_i = \sqrt{N_{\rm meas}}$. 
The current (expected) bound for the resolved signature, with these simplifications, is shown as solid (dashed) orange line in \cref{Tutorial:ALPphotonATeecolliders:BelleII:Bounds}. Note that, as expected, the bound starts at ALP masses of a few hundred MeV, and worsens as the $m_a$ increases and reaches the boundary of allowed kinematics, in which case the production cross section quickly drops to zero, see \cref{eq:Tutorial:ALP-photon:ProdCrossSec:AssocProd}.

\clearpage

\bibliographystyle{JHEP}
\bibliography{references}

\end{document}